\documentclass{jpp}

\usepackage{graphicx}
\usepackage{natbib}

\ifCUPmtlplainloaded \else
  \checkfont{eurm10}
  \iffontfound
    \IfFileExists{upmath.sty}
      {\typeout{^^JFound AMS Euler Roman fonts on the system,
                   using the 'upmath' package.^^J}%
       \usepackage{upmath}}
      {\typeout{^^JFound AMS Euler Roman fonts on the system, but you
                   dont seem to have the}%
       \typeout{'upmath' package installed. JPP.cls can take advantage
                 of these fonts, if you use 'upmath' package.^^J}%
      }
  \else
  \fi
\fi

\ifCUPmtlplainloaded \else
  \checkfont{msam10}
  \iffontfound
    \IfFileExists{amssymb.sty}
      {\typeout{^^JFound AMS Symbol fonts on the system, using the
                'amssymb' package.^^J}%
       \usepackage{amssymb}%
         \let\leq=\leqslant
         
      }{}
  \fi
\fi

\ifCUPmtlplainloaded \else
  \IfFileExists{amsbsy.sty}
    {\typeout{^^JFound the 'amsbsy' package on the system, using it.^^J}%
     \usepackage{amsbsy}}
    {\providecommand\boldsymbol[1]{\mbox{\boldmath $##1$}}}
\fi

\newsavebox{\astrutbox}
\sbox{\astrutbox}{\rule[-5pt]{0pt}{20pt}}

\title[Kinetic linear theory]{Comparison of linear modes in kinetic plasma models}

\author[E. Camporeale and D. Burgess]%
{Enrico Camporeale$^1$%
  \thanks{Email address for correspondence: e.camporeale@cwi.nl},\ns
\and David Burgess$^2$}

\affiliation{$^1$Center for Mathematics and Computer Science (CWI), Amsterdam, The Netherlands\\[\affilskip]
$^2$School of Physics and Astronomy, Queen Mary University of London, London, UK}

\pubyear{2016}
\volume{}
\pagerange{}
\date{?; revised ?; accepted ?. - To be entered by editorial office}
\begin{document}

\maketitle

\begin{abstract}
We compare, in an extensive and systematic way, linear theory results obtained with the hybrid (ion-kinetic and electron-fluid), the gyrokinetic and the fully-kinetic plasma models. We present a test case with parameters that are relevant for solar wind turbulence at small scales, which is a topic now recognized to need a kinetic treatment, to a certain extent.
We comment on the comparison of low-frequency single modes (Alfv\'{e}n/ion-cyclotron, ion-acoustic, and fast modes) for a wide range of propagation angles, and on the overall spectral properties of the linear operators, for quasi-perpendicular propagation. The methodology and the results presented in this paper will be valuable when choosing which model should be used in regimes where the assumptions of each model are not trivially satisfied.
\end{abstract}

\begin{PACS}
\end{PACS}

\section{Introduction}\label{sec:introduction}
One of the distinctive characteristics of plasmas is that they have time and length scales that are typically well separated.
For example, in a electron-proton plasma the mass ratio between the two species is approximately equal to 1836, and this determines a separation of the same factor
in the respective cyclotron frequencies, and of a factor $\sim\sqrt{1836}\simeq 43$ in the respective Larmor radii and plasma frequencies. 
Such a large separation of scales translates very often into unaffordable costs, from a computational point of view. In particular, 
it is well known that a fully-kinetic treatment, where two Vlasov equations (one for each species) are coupled to the Maxwell's equations, is prohibitively expensive in multiple dimensions. 
For this reason, several reduced models have been studied in the past, that are able to retain some of the kinetic features of the plasma, at 
a reduced computational cost \citep{matthews94, waltz97, park99, cheng99, goswami05, brizard07, valentini07}. Such models could be ordered in a hierarchy that transitions all the way from kinetic to fluid theory \citep{dendy95, tronci14, tronci15, camporeale16}. 
In this paper, we focus specifically on two models that have been extensively used for solar wind turbulence studies: the hybrid (ion-kinetic and electron-fluid) and the gyrokinetic models. The philosophy behind these two models is quite different. 
In the hybrid model the heavy species (protons) is treated kinetically and its distribution function obeys the Vlasov equation (which is in principle six-dimensional), while the light species (electrons) is treated as an inertia-less neutralizing fluid. Thus, the computational saving is achieved by factoring out from the model the time and spatial scales associated with electrons. We note that the term "hybrid" is used widely in different contexts with different meanings, but here we use it with the specific meaning given.
On the other hand, the gyrokinetic model treats both species kinetically, but it performs a time average on the ion cyclotron period. As a result, the six-dimensional distribution functions collapse into five dimensions. Interestingly, such dimensionality reduction of the particle distribution functions (and the fact that the dynamics faster than the ion gyro-period are neglected) results into a sizable saving in computational time.\\
Like all reduced models, the derivation of the hybrid and gyrokinetic models is based on a certain number of assumptions. They both make use of the so-called quasi-neutrality condition, which asserts that ion and electron densities are equal, $n_i=n_e$. This is satisfied as long as the Debye length is much smaller than the characteristic spatial scales, and the characteristic frequencies are much smaller than the electron plasma frequency. As a consequence, the displacement current in Ampere's law is neglected in both models. Furthermore, the gyrokinetic theory assumes that frequencies are much smaller than the ion cyclotron frequency: $\omega\ll\Omega_i$, and that the wave spectrum is very anisotropic: $k_{||}\ll k_\perp$. The hybrid model, instead, necessitates of a closure in terms of an equation of state for the electrons.\\
The fundamental problem that we tackle in this paper is: what does \emph{much smaller} mean?\\
Once a model (and the computer code that solves it) has been extensively used and benchmarked in a certain regime, it is certainly tempting to push the simulations into unexplored regimes. Although there are cases for which the reduced models are obviously inadequate, there exist a large range of 'gray areas', where their validity is not obvious. In a sense, the issue of validating a reduced model is an ill-posed problem. In principle, the validation of the results in these gray areas can only be performed by comparing with a fully-kinetic model. On the other hand, if such solutions from the fully-kinetic models were available, one would not need to use a reduced model!\\
The rationale for this work stems from the fact that recent simulations performed with hybrid and the gyrokinetic codes might be considered the current state-of-the-art in studies concerned with solar wind turbulence (see, e.g., \citep{howes11, valentini10, servidio12, franci15a}). 
Indeed, it is now widely recognized that a kinetic treatment is necessary for studying solar wind turbulence, at small scales \citep{camporeale11, sahraoui12, boldyrev13, haynes14, vasconez14, haynes15}.
Therefore, it is legitimate to ask what is the range of validity of such models, and when the results start to deviate considerably from the correct solutions. Here, we focus on comparing solutions obtained within the linear theory. Clearly, the ability of a model to describe the property of linear waves, within a certain accuracy, can be interpreted as a necessary (but not sufficient) condition for the model to yield reliable results in the non-linear regime.\\
The scope of this paper is twofold. First, we directly compare linear theory solutions obtained from the hybrid and the gyrokinetic models with the complete Vlasov-Maxwell model, for wavevectors $k$ ranging from $k\rho_i=0.1$ to $k\rho_i=44$ (with $\rho_i$ the ion Larmor radius), and for a wide range of propagation angles $\theta$, ranging from $10^\circ$ to $85^\circ$; second, we present a robust and novel methodology to perform such comparison in a fair way, that takes into account the fact that for large wavevector and/or propagation angles obtaining linear solutions  is a non-trivial numerical exercise, due to the presence of many normal modes.\\
The paper is organized as follows. We introduce the methodology and the parameters chosen for a test case in Section \ref{sec:methodology}, where we also define the errors that will be measured. Section \ref{sec:results} shows the comparison between hybrid, gyrokinetic, and fully-kinetic solutions for a wide range of cases. Conclusions are drawn in Section \ref{sec:conclusions}.

\section{Methodology and chosen parameters}\label{sec:methodology}
We study a plasma composed of electrons and protons, with mass ratio set to the physical value $m_i/m_e=1836$, in Cartesian coordinates $(\hat{i},\hat{j},\hat{k})$. As it is customary in linear theory, we study an uniform Maxwellian plasma with density $n_0$, in a constant magnetic field $\mathbf{B}=B_0\hat{k}$, that is perturbed with a small-amplitude plane wave of the form $\exp[i(\mathbf{k}\cdot \mathbf{x} -\omega t)]$, with wavevector $\mathbf{k}=(k_\perp,0,k_{||})$, and complex frequency $\omega=\omega_r + i\gamma$.\\
We focus on a single choice of parameters that are relevant for solar wind turbulence studies. In order to facilitate comparisons with published literature, we use the typical solar wind parameters studied in \citet{salem12}: $B_0 = 11$ nT, $T_e = 13.0$ eV (electron temperature), $T_p = 13.6$ eV (proton temperature), $n_0 = 9 $ cm$^{-3}$. The proton plasma beta is $\beta_p =0.4$, and the electron plasma to cyclotron frequency ratio is $\omega_{pe}/\Omega_e \simeq 88$. The ratio between Alfv\'{e}n and ion thermal velocities is $v_A/v_{ti}=1.57$, and the ratio between Alfv\'{e}n velocity and speed of light is $v_A/c=2.6\times10^{-4}$.\\
In all of the three cases considered (fully-kinetic, hybrid, and gyrokinetic), the linear wave solutions are obtained by finding the complex frequencies $\omega$ for which the determinant of a $3\times 3$ matrix $\mathbf{D}$ vanishes. $f(\omega,\mathbf{k})=\det(\mathbf{D})$ is a nonlinear function of both $\mathbf{k}$ and $\omega$, and the standard way to numerically solve the problem, for a fixed $\mathbf{k}$, is by employing an iterative root finder based on Newton's method. Unfortunately, although it is straightforward to extend Newton's method for finding zeros in the complex plane, the method can be very sensitive to the choice of an initial guess. Moreover, due to the possible presence of multiple solutions which are close to each other in the complex plane, it is not always easy to track a solution $\omega$ varying $\mathbf{k}$. In other words, the solver can easily jump from a solution branch to another, whenever two solutions get close enough. The paper by \citet{podesta12} explains how to reduce the numerical error in calculating the elements of $\mathbf{D}$ in the case of the Vlasov-Maxwell system. \citet{podesta12} also suggests identifying possible roots by visual inspection of the function $1/|f|$, where the peaks might indicate roots. An alternative way, which we think is more robust, less time consuming, and can be easily automated, is to analyze separately the real $(\Re)$ and imaginary $(\Im)$ parts of the function $f(\omega)$, and to identify the curves in the complex plane, where $\Re(f)=0$ and $\Im(f)=0$, that is where they change sign. Clearly, the intersections of such curves ($\Re(f)=\Im(f)=0$) are the roots $f(\omega)=0$, that can be used as initial guess for the Newton solver, to achieve a better accuracy. We have employed this method to identify \emph{all} the roots in a given region of the complex plane, in Section \ref{sec:def_err}\\
We refer to \citet{howes06} for a derivation of the gyrokinetic linear theory. The linear theory for the hybrid model can easily be derived by noting that, in the standard derivation of the Vlasov-Maxwell linear theory, the dielectric tensor $\varepsilon(\omega,\mathbf{k})$ is additive by species (see, e.g. \citet{stix62}):
\begin{equation}
\varepsilon(\omega,\bf{k}) = \bf{1} + \sum_s\chi_s(\omega,\bf{k}),
\end{equation}
where $\mathbf{1}$ is the unit dyadic, $\chi_s$ is the susceptibility of species $s$, and the sum is over plasma species.
Therefore, in a hybrid model it is sufficient to calculate the electron susceptibilities $\chi_e$ derived within a two-fluid linear model. Here, we have employed the so-called warm plasma dispersion relation derived in \citet{swanson12}, which assumes the equation of state $p_e/n_e = v_{te}^2$ (with $p_e$, $n_e$, and $v_{te}$ the electron pressure, density, and thermal velocity, respectively). Moreover, in order to enforce the quasi-neutrality condition, it sufficient to suppress the unit dyadic term in the dielectric tensor, which is thus defined simply as the sum of the ion susceptibility (derived in kinetic theory) and the electron one (derived in two-fluid theory). 

\subsection{Diagnostics}\label{sec:diagnostics}
\begin{figure}
  \includegraphics[width=10cm]{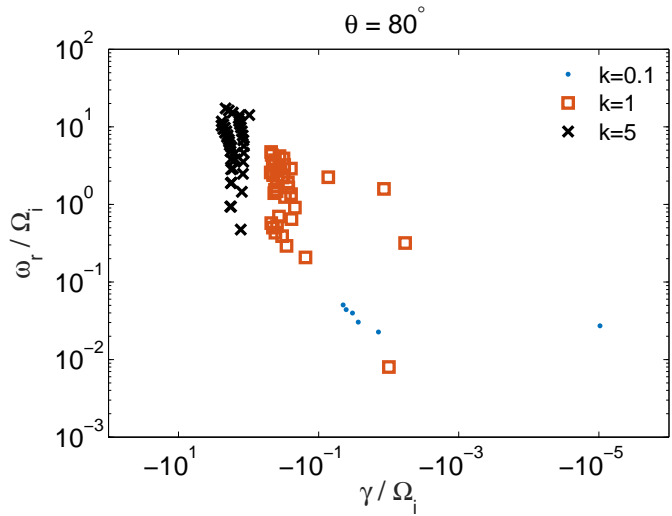}
  \caption{Normal modes derived with the fully-kinetic model. Damping rate ($\gamma$) and real frequency $(\omega_r)$ are on horizontal and vertical axis, respectively. The angle of propagation is $\theta=80^\circ$. Blue dots, red circles, and black square denote $k\rho_i=0.1$, 1, 5, respectively.}
\label{fig:modes_80}
\end{figure}

\begin{figure}
  \includegraphics[width=10cm]{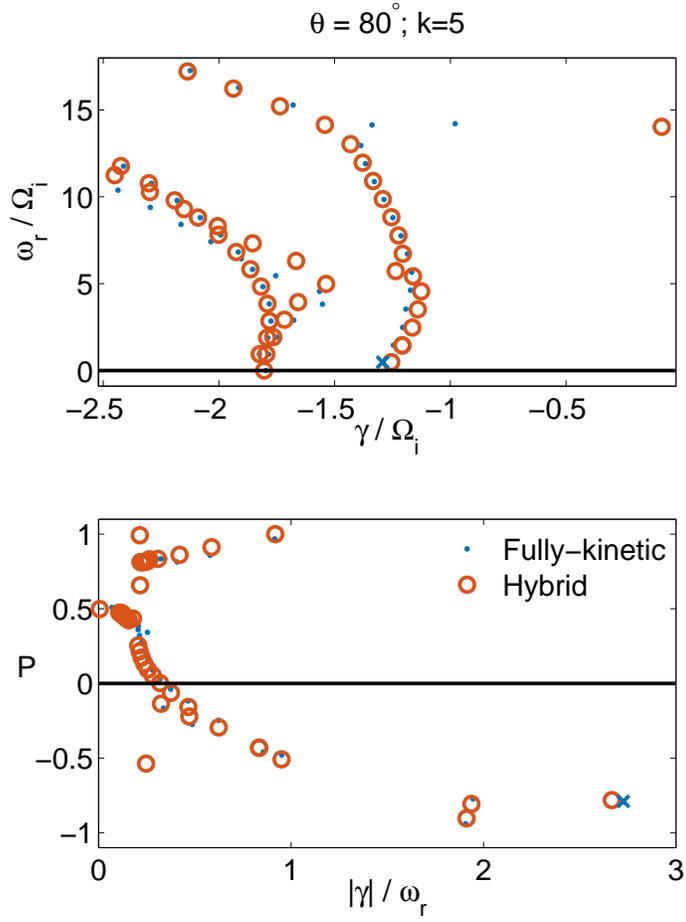}
  \caption{Normal modes for $\theta=80^\circ$ and $k\rho_i=5$ for fully-kinetic (blue dots) and hybrid (red circles).
  Top panel: solutions in the complex plane $(\gamma,\omega_r)$. Bottom panel: solutions plotted as functions of $|\gamma|/\omega_r$ (horizontal axis) and polarization P (vertical axis). In both panels the Kinetic Alfv\'{e}n Wave (KAW) is denoted with a blue cross.}
\label{fig:modes_5_80}
\end{figure}

The results obtained from hybrid and gyrokinetic linear theory are compared with the ones obtained in the fully-kinetic Vlasov-Maxwell theory. 
We divide the parameter space into \emph{oblique} propagation, for $\theta\leq 60^\circ$, and \emph{quasi-perpendicular} propagation, for $\theta>60^\circ$.
While the diagnostics used for oblique propagation are based on the comparison of single normal modes, we argue that such diagnostics are not sufficiently informative, for the quasi-perpendicular regime. Typically, among the infinite number of modes which can in principle exist as solutions of the full linear Vlasov-Maxwell system, one is interested in the few that have lower damping rates, while the others (usually referred to as 'heavily-damped') are of little interest because, supposedly, they do not play a significant role for mode coupling.
We note that this view, although probably correct when the modes are well separated, is questionable when there are a multitude of normal modes clustered together. In this case, one would like to take in considerations the overall spectral properties of the linear operator. Such approach is justified by the \emph{non-normality} of the Vlasov-Maxwell linear operator. We refer to, e.g., \citep{camporeale09, camporeale10, podesta10, ratushnaya14, friedman14} and citations therein, for a discussion on  non-normal linear operators in plasma physics. In short, and for the purpose of this paper, we recall that non-normality is due to the non-orthogonality of the eigenvectors of a linear operator, and is connected to the phenomena of transient growth, by-pass transitions and, generally, to mode coupling. An important characteristic of non-normal operators is that their eigenvalues are highly sensitive to perturbations, in the sense that the spectrum can be easily distorted in the complex plane, when the original operator is slightly perturbed. This concept gives rise to the idea of pseudospectrum, which is extensively discussed in \citet{trefethen05} and, for the case of the Vlasov-Maxwell system, in \citet{camporeale12}.
For instance, Figure \ref{fig:modes_80} shows a subset of the normal modes (obtained in the fully-kinetic model), in the complex plane, for $\theta=80^\circ$. The damping rate $\gamma$ and the real frequency $\omega_r$ are on horizontal and vertical axis, respectively (here and in all figures they are normalized to the ion cyclotron frequency $\Omega_i$). The different symbols are for $k\rho_i=0.1$ (blue dots), $k\rho_i=1$ (red circles), and $k\rho_i=5$ (black crosses). 
Clearly, a distinction between modes purely based on their damping rate becomes less and less possible with increasing wavevector. 
Moreover, there is no reason to attribute a particular importance to some well-characterized modes (that is, well-characterized in the low $k$ and oblique propagation regime), such as the  Kinetic Alfv\'{e}n Wave (KAW) mode. This is particularly clear by inspection of Figure \ref{fig:modes_5_80}.
Here, we plot, in the top panel, the subset of normal modes that are confined in the complex plane within the domain $-2.5<\gamma<0$ and $0<\omega_r<20$.
The blue dots and red circles denote the fully-kinetic and the hybrid solutions, respectively. The Kinetic Alfv\'{e}n wave, that is the root of the dispersion relation that when tracked to low $k$ and low $\theta$ is connected to the Alfv\'{e}n wave mode, is denoted with a blue cross ($\omega_r = 0.47$, $\gamma=-1.29$). It is evident that, in this regime, KAW is surrounded by many other modes, some of them with similar frequency but lower damping rate. Nevertheless, KAW has been attributed a predominant role in some of the recent solar wind literature \citep{sahraoui09, salem12}. Even if one would like to adhere to the classical view where the least damped modes are more important than the heavily-damped one, it is clear that KAW is not special. The bottom panel of Figure \ref{fig:modes_5_80} shows the same modes, but now sorted as function of their damping rate per wave period $(|\gamma|/\omega_r)$ (horizontal axis), and their polarization $P$ (vertical axis). The polarization is here defined following \citet{hunana13}, as $P=\arg(E_y/E_x)/\pi$, with $E_y$ and $E_x$ complex amplitudes of the electric field. Hence, negative/positive values of $P$ represent left/right hand polarization. Interestingly, in this regime, KAW is the most rapidly damped mode per wave period. A final remark on Figure \ref{fig:modes_5_80} is that one can see that, despite being in the electron range (i.e. $k\rho_i>1$), the hybrid model captures most of the normal modes quite accurately. It creates a spurious mode with $\omega_r=14$ and $\gamma=-0.08$ due to the inability to account for electron Landau damping. 

\subsubsection{Definition of errors}\label{sec:def_err}
\begin{figure}
  \includegraphics[width=15cm]{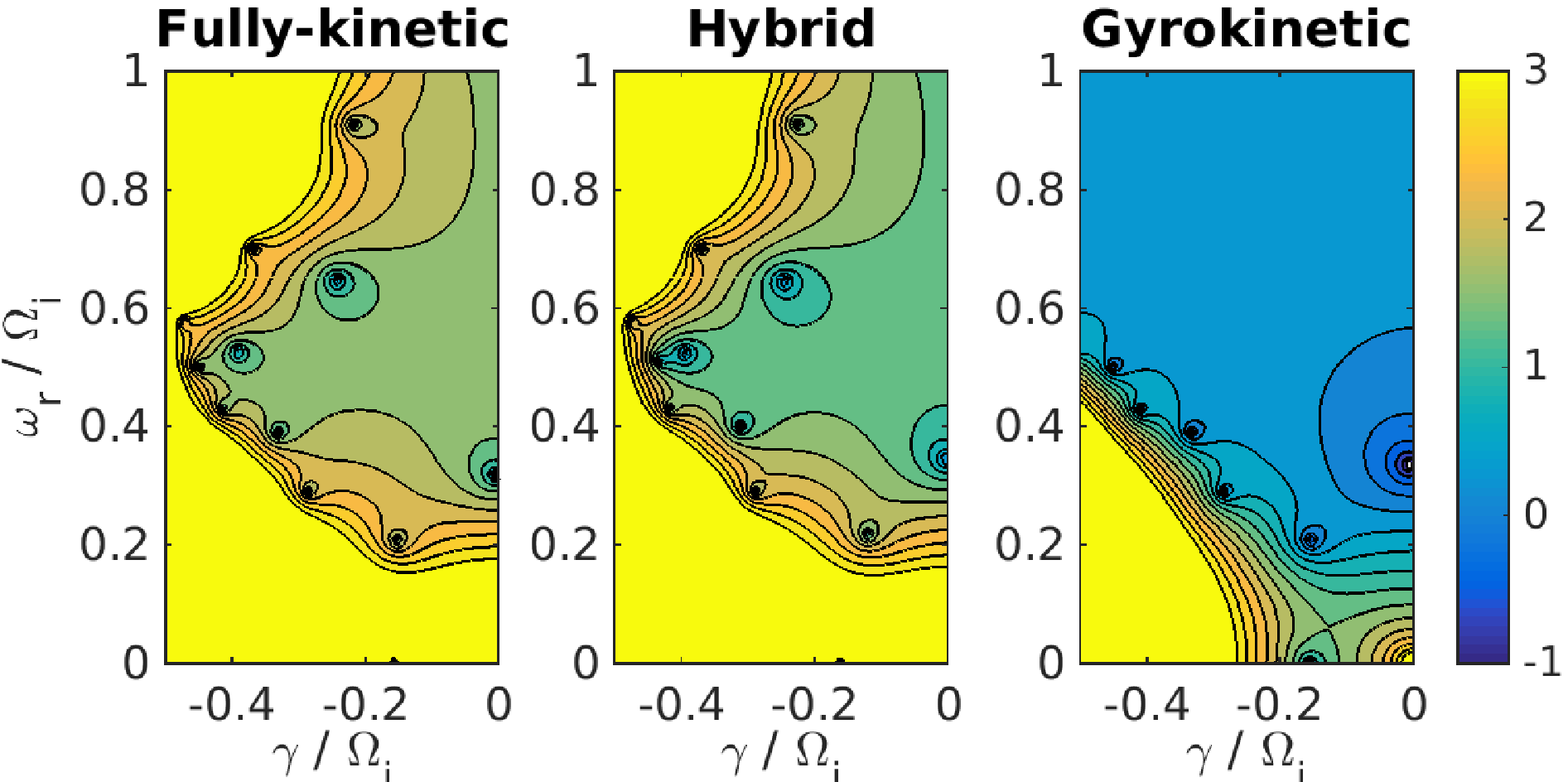}
  \caption{Pseudospectra in the complex plane $(\gamma,\omega_r)$ for $k\rho_i = 1$, and $\theta = 80^\circ$. Left: full-kinetic; middle: hybrid; right: gyrokinetic. The color plot represents $f(\omega)=\det(\bf{D})$ in logarithmic scale.}
\label{fig:pseudo}
\end{figure}

\begin{figure}
  \includegraphics[width=15cm]{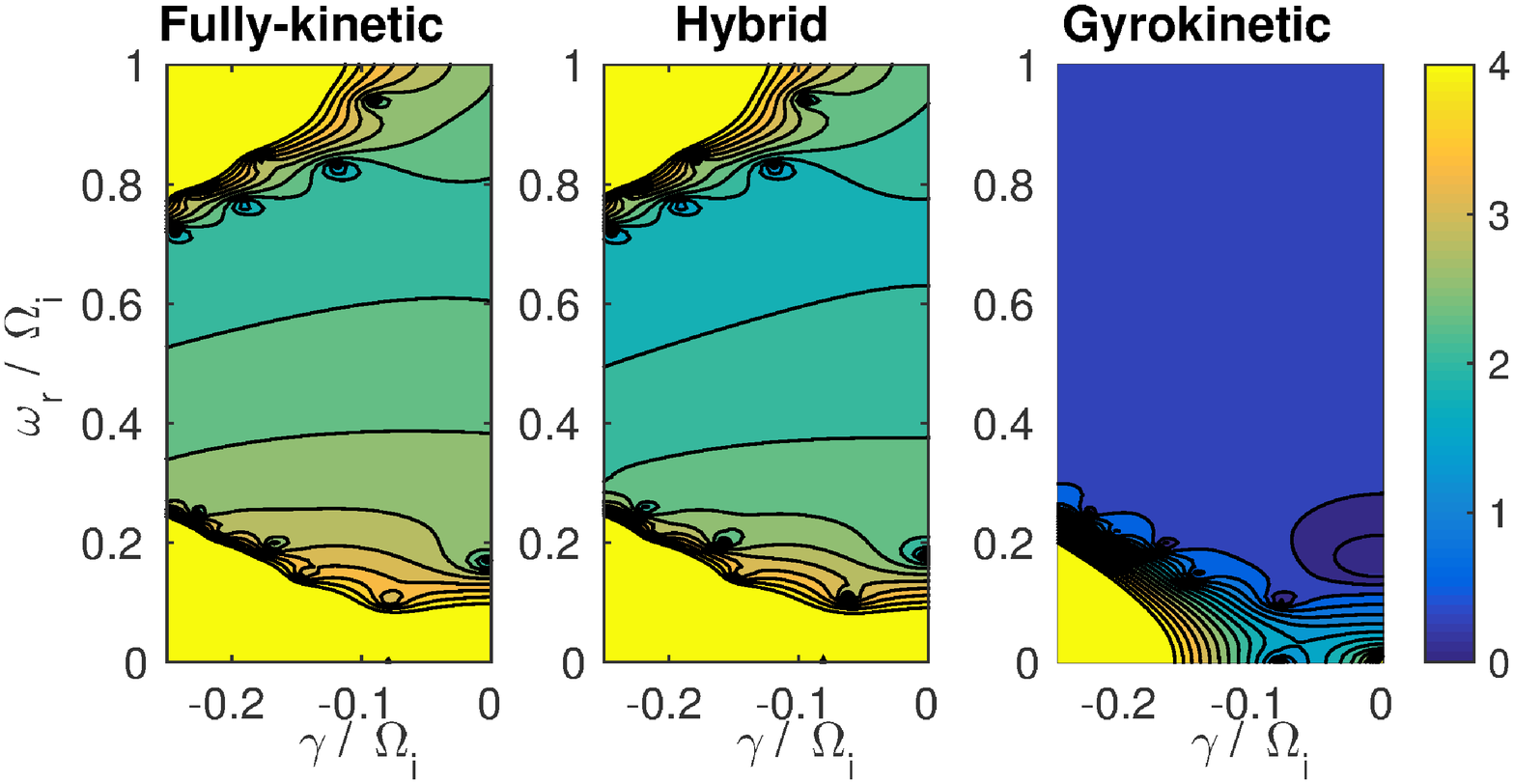}
  \caption{Pseudospectra in the complex plane $(\gamma,\omega_r)$ for $k\rho_i = 1$, and $\theta = 85^\circ$. Left: full-kinetic; middle: hybrid; right: gyrokinetic. The color plot represents $f(\omega)=\det(\bf{D})$ in logarithmic scale.}
\label{fig:pseudo_k_1_85}
\end{figure}

We define the following two measures of error, that assess the deviation of the frequency $\omega$ (obtained either with hybrid or gyrokinetic), with respect to the correct value obtained with the Vlasov-Maxwell theory, denoted as $\omega_{VM}$:
\begin{equation}
 \varepsilon_\omega = \frac{|\omega_{VM}-\omega|}{|\omega_{VM}|}\label{err_1}
\end{equation}
\begin{equation}
 \varepsilon_{wave} = ||\Re\left({\exp[-i\omega_{VM} t]}\right) - \Re\left({\exp[-i\omega t]}\right) ||_2\label{err_2}
\end{equation}
The error $\varepsilon_\omega$ in Eq.(\ref{err_1}) is simply the relative distance between the two frequencies, in the complex plane, and it does not distinguish the measure in which the real and the imaginary part (damping rate) of the frequency contribute to the error.
In order to calculate the error denoted as $\varepsilon_{wave}$ in Eq.(\ref{err_2}), we define a time interval $t=[0, 2\pi/\Re(\omega_{VM})]$, that covers one wave period (with respect to $\omega_{VM}$), and is discretized in 200 points, and we calculate the $L_2$ norm of the difference of the real part of the wave amplitude (normalized by the number of points). This metric is relevant because it comprises the information on how much a wave with frequency $\omega$ will differ, during one wave period, from the reference wave with frequency $\omega_{VM}$, by modeling its actual time evolution.\\
Moreover, we will use one more metric to compare results in the quasi-perpendicular regime.
Here, we seek to have a measure that gives us information on the global difference between the spectra calculated with the three methods. 
For this purpose, we rely on the (generalized) definition of the $\varepsilon$-pseudospectrum $\Lambda_\varepsilon(\mathbf{D})$, for a fixed $\mathbf{k}$:
\begin{equation}
 \Lambda_\varepsilon(\mathbf{D})=\{z\in\mathbb {C}: |\det(\boldsymbol{D}(z))| \leq \varepsilon \}.
\end{equation}
Clearly, the pseudospectrum reduces to the standard spectrum (set of eigenvalues) for $\varepsilon\rightarrow 0$. 
For instance, Figures \ref{fig:pseudo} and \ref{fig:pseudo_k_1_85} compare the fully-kinetic (left), hybrid (middle), and gyrokinetic (right) pseudospectra for the case $k\rho_i=1$ and $\theta=80^\circ$ (Figure \ref{fig:pseudo}) or $\theta=85^\circ$ (Figure \ref{fig:pseudo_k_1_85}), in the region $-0.5<\gamma<0$, and $0<\omega_r<1$. The color plot is in logarithmic scale and denotes the value of $\det(\mathbf{D})$. 
One can visually identify the eigenvalues as the points where the concentric curves are centered.
In order to measure quantitatively the error in the pseudospectra, we proceed as follows. 
We compute, for a fixed value of $k=|\mathbf{k}|$ the function $f(\omega)=\det(\mathbf{D(\omega)})$ on a $500\times 500$ grid covering the domain
$0\leq\omega_r\leq5k$, and $-0.5k\leq\gamma\leq0$. Because $f$ can vary several orders of magnitude within this domain, it is not convenient to calculate an error directly defined on $f$. Moreover, we seek to define an error that emphasizes the differences only in regions where $f(\omega_{VM})$ is small, disregarding regions where $f$ is large. Therefore, we map $f(\omega)$ into a new function $\phi(\omega)$ through a sigmoid:
\begin{equation}
 \phi(\omega) = 1+ \frac{9f}{1+f}.
\end{equation}
The new function $\phi(\omega)$ is defined between 1 (for $f=0$) and 10 (for $f\rightarrow\infty$).
Finally, we define the error $\varepsilon_D$ as:
\begin{equation}
 \varepsilon_D = \left\|\frac{ (\phi(\omega) - \phi(\omega_{VM}))}{\phi(\omega_{VM})}\right\|_2,\label{eps_D}
\end{equation}
where the L-2 norm is normalized on the number of grid points. 

\subsection{Range of validity of the gyrokinetic model}
\begin{figure}
  \includegraphics[width=10cm]{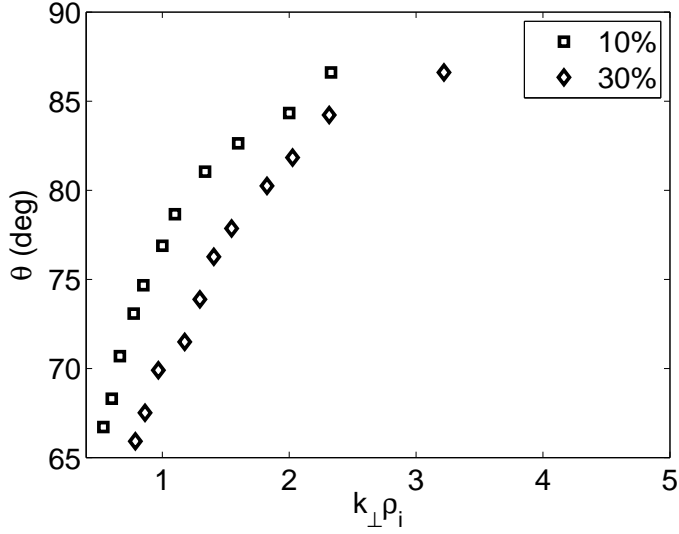}
  \caption{Range of validity of the gyrokinetic model. Square and diamond symbols indicate an error in the real frequency of 10\% and 30\%, respectively.
  The perpendicular wavevector $k_\perp \rho_i$ and the angle of propagation $\theta$ are on horizontal and vertical axis. The model is valid above the symbols. }
\label{fig:validity_GK}
\end{figure}

\begin{figure}
  \includegraphics[width=10cm]{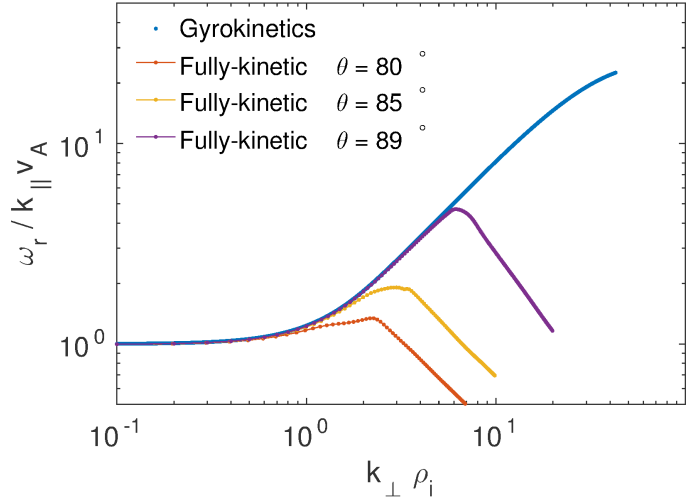}
  \caption{Range of validity of the gyrokinetic model. Dispersion relation for Alfv\'{e}n/ion-cyclotron wave, plotted as $\omega_r/k_{||}v_A$ as a function of $k_\perp\rho_i$. The blue line denotes the gyrokinetics solution, which is independent of the angle of propagation. The red, yellow, and purple lines denote the fully-kinetic solution for angles of propagation $\theta = 80^\circ,85^\circ,89^\circ$. }
\label{fig:validity_GK_2}
\end{figure}

As we mentioned, one important assumption in the derivation of the gyrokinetic model is that $k_{||}\ll k_\perp$. This requirement restricts dramatically the range of validity of the model. Following the philosophy outlined in the Introduction, we will compare the three models  
also in the non quasi-perpendicular propagation regime, that is, in a range where the gyrokinetic model is not expected to be valid.
Figure \ref{fig:validity_GK} shows the range of validity of the gyrokinetic model. The symbols denote the angle of propagation $\theta$ for a given value of perpendicular wavevector $k_\perp\rho_i$, that results in a relative error on the Kinetic Alfv\'{e}n Wave real frequency of $10\%$ (squares) and $30\%$ (diamonds). One can appreciate that for $k_\perp\rho_i>3$ the propagation cone is restricted to angles very close to $90^\circ$. 
An alternative way of appreciating this result is shown in Figure \ref{fig:validity_GK_2}, that shows the dispersion relation for the Alfv\'{e}n/ion-cyclotron wave. When plotted as $\omega_r/k_{||}v_A$ as a function of $k_\perp\rho_i$, the dispersion relation becomes independent of angle, in gyrokinetics (shown in blue). The red, yellow, and purple lines denote the fully-kinetic solution for angles of propagation $\theta = 80^\circ,85^\circ,89^\circ$. 
The gyrokinetic model has been suggested as a valid model for the study of solar wind turbulence all the way to electron scales \citep{howes11}, on the basis that some observational data suggest that the propagation angle in the solar wind averages at about 88 degrees, and across the examined spectra never fall below 60 degrees \citep{sahraoui10, narita10}.
It is not within the scope of this paper to discuss the issue of the observed wave spectrum anisotropy in the solar wind. However, it is fair to mention that the problem is still open, and some recent works have emphasized the importance of quasi-parallel wave propagation \citep{osman09, kiyani13, lion16}. Furthermore, the development of wave vector anisotropy is an open research question, for which linear properties over a range of propagation angles will be important. It is in this spirit that, in the following, we present a comparison of the models across a wide range of propagation angles.

\section{Results}\label{sec:results}
Historically, different communities in plasma physics have adopted different nomenclatures for linear waves. Here we follow the nomenclature adopted by \citet{gary_book}. We will refer to Alfv\'{e}n/ion-cyclotron as the electromagnetic mode that is left-handed polarized for parallel propagation. The ion-acoustic mode is the one that connects to the electrostatic solution for parallel propagation, and the fast (magnetosonic) waves is the mode that connects to the right-handed polarized mode for parallel propagation.
In this section we first discuss the dispersion relations of each one of these three modes, and then we compare the models comparing the pseudo-spectra in the quasi-perpendicular regime. 

\subsection{Alfv\'{e}n/ion-cyclotron (AIC)}
\begin{figure}
  \includegraphics[width=10cm]{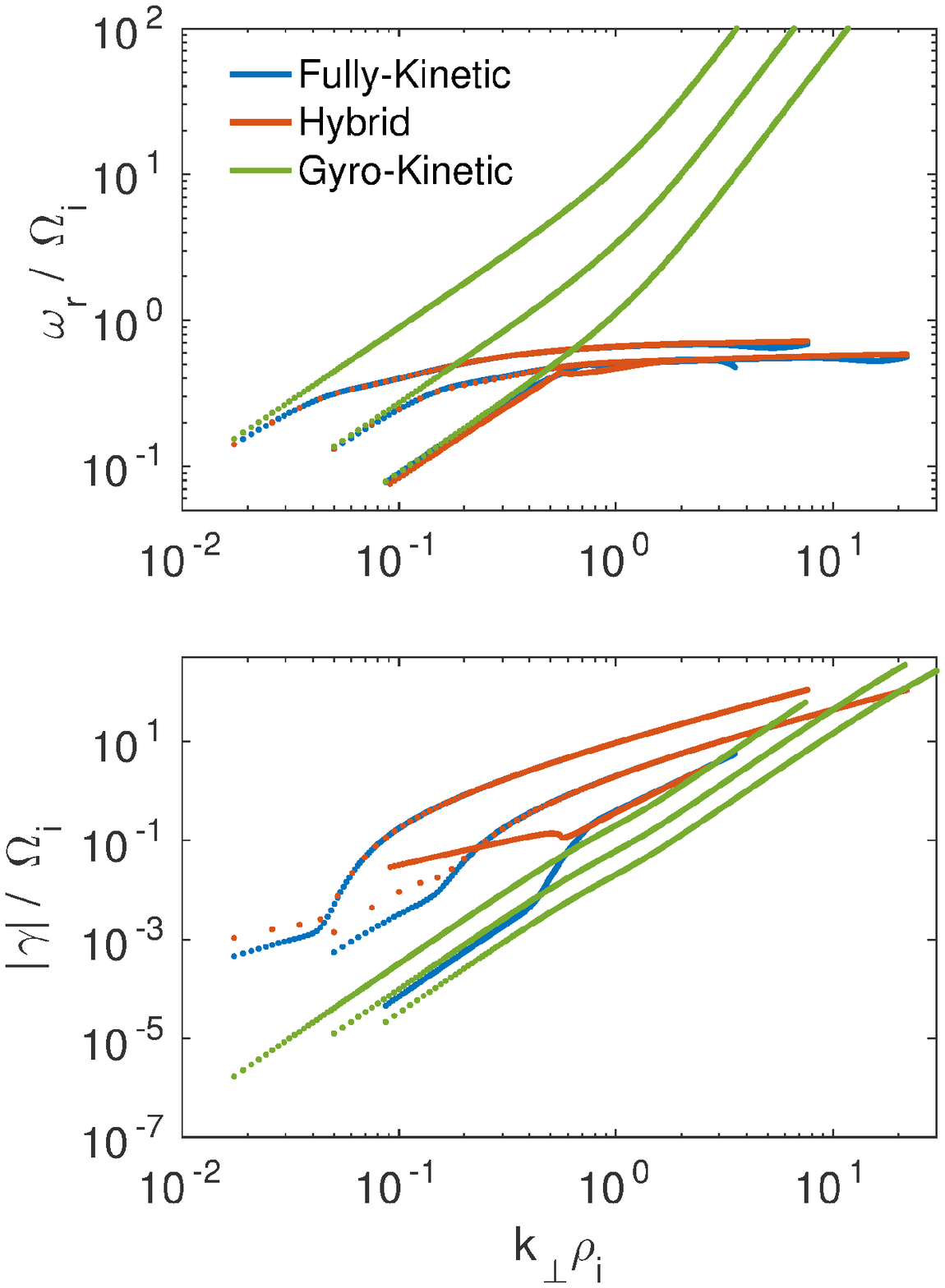}
  \caption{Dispersion relation for Alfv\'{e}n/ion-cyclotron waves for angles of propagation $\theta=10^\circ, 30^\circ, 60^\circ$ (curves from left to right). Blue, red, and green lines are for fully-kinetic, hybrid, and gyrokinetic, respectively. Top panel: real frequency $\omega_r$; bottom panel: damping rate $\gamma$. $k_\perp\rho_i$ is on the horizontal axis. All axis are in logarithmic scale.}
\label{fig:disp_rel_AIC}
\end{figure}

\begin{figure}
  \centerline{\includegraphics[width=10cm]{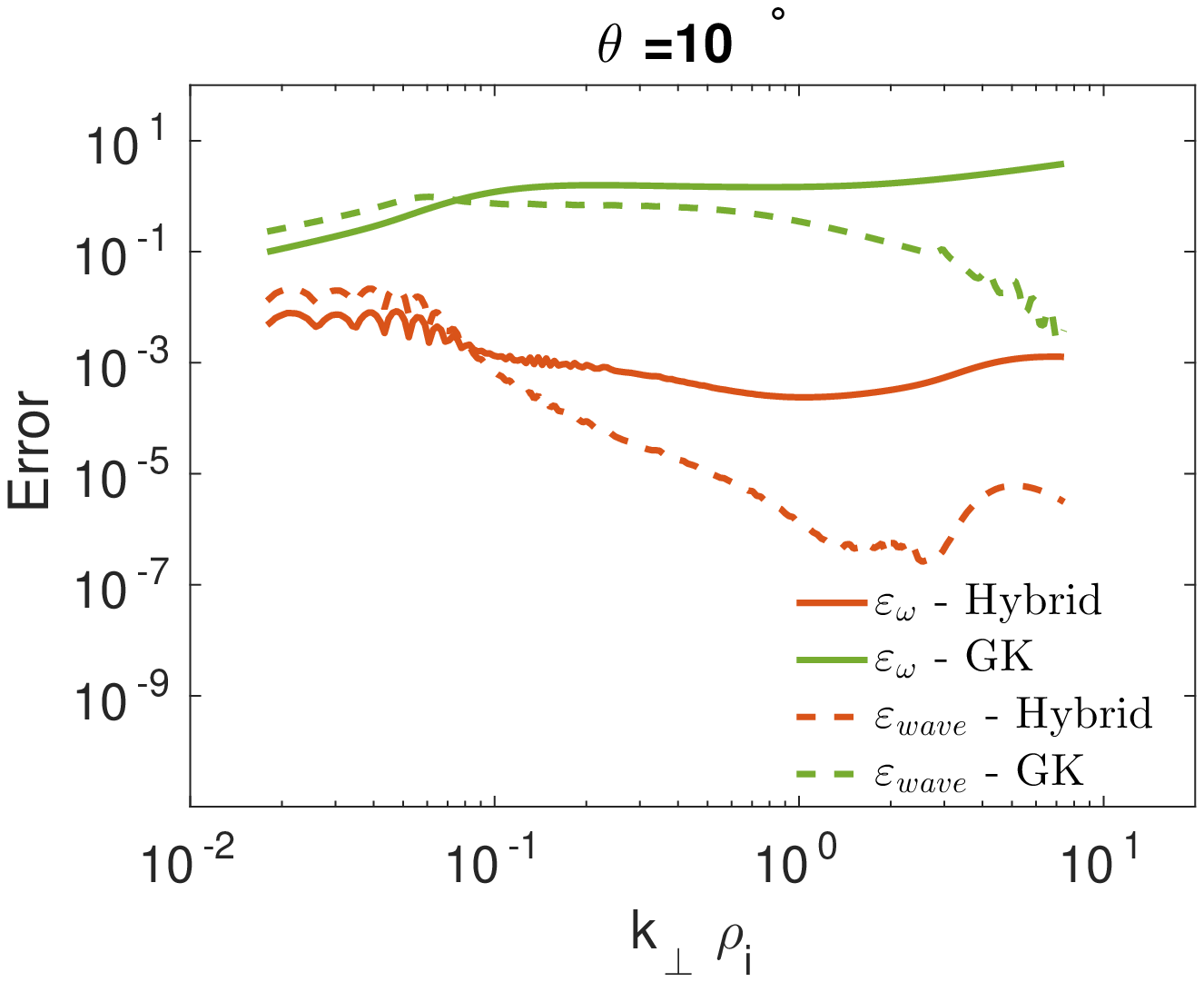}}
  \caption{Errors for Alfv\'{e}n/ion-cyclotron wave with $\theta=10^\circ$, as functions of $k_\perp\rho_i$. Red and green curves are for hybrid and gyrokinetics, respectively. Solid lines denote $\varepsilon_\omega$, and dashed lines denote $\varepsilon_{wave}$.}
\label{fig:error_AIC_10}
\end{figure}

\begin{figure}
  \centerline{\includegraphics[width=10cm]{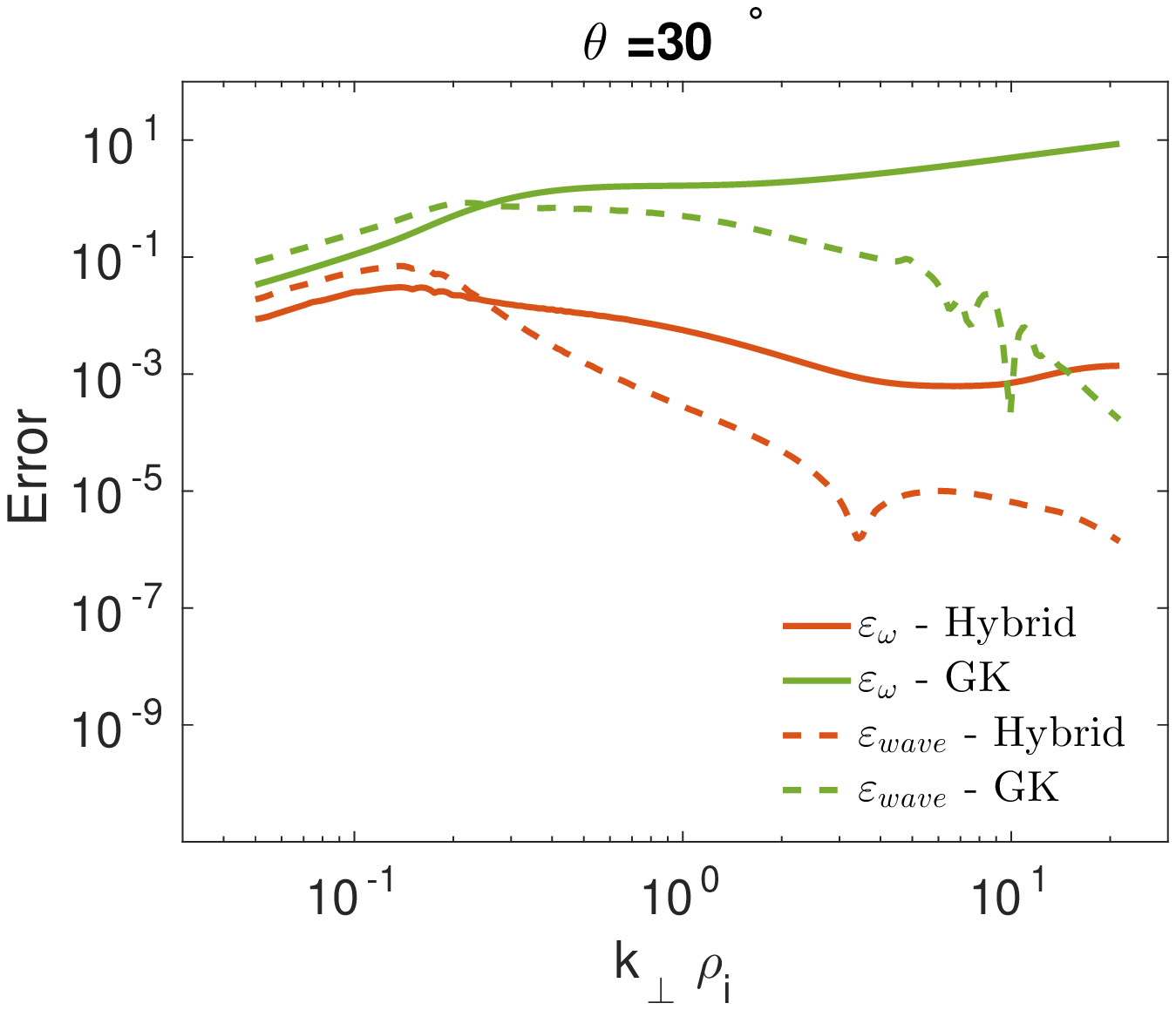}}
  \caption{Errors for Alfv\'{e}n/ion-cyclotron wave with $\theta=30^\circ$, as functions of $k_\perp\rho_i$. Red and green curves are for hybrid and gyrokinetics, respectively. Solid lines denote $\varepsilon_\omega$, and dashed lines denote $\varepsilon_{wave}$.}
\label{fig:error_AIC_30}
\end{figure}

\begin{figure}
  \centerline{\includegraphics[width=10cm]{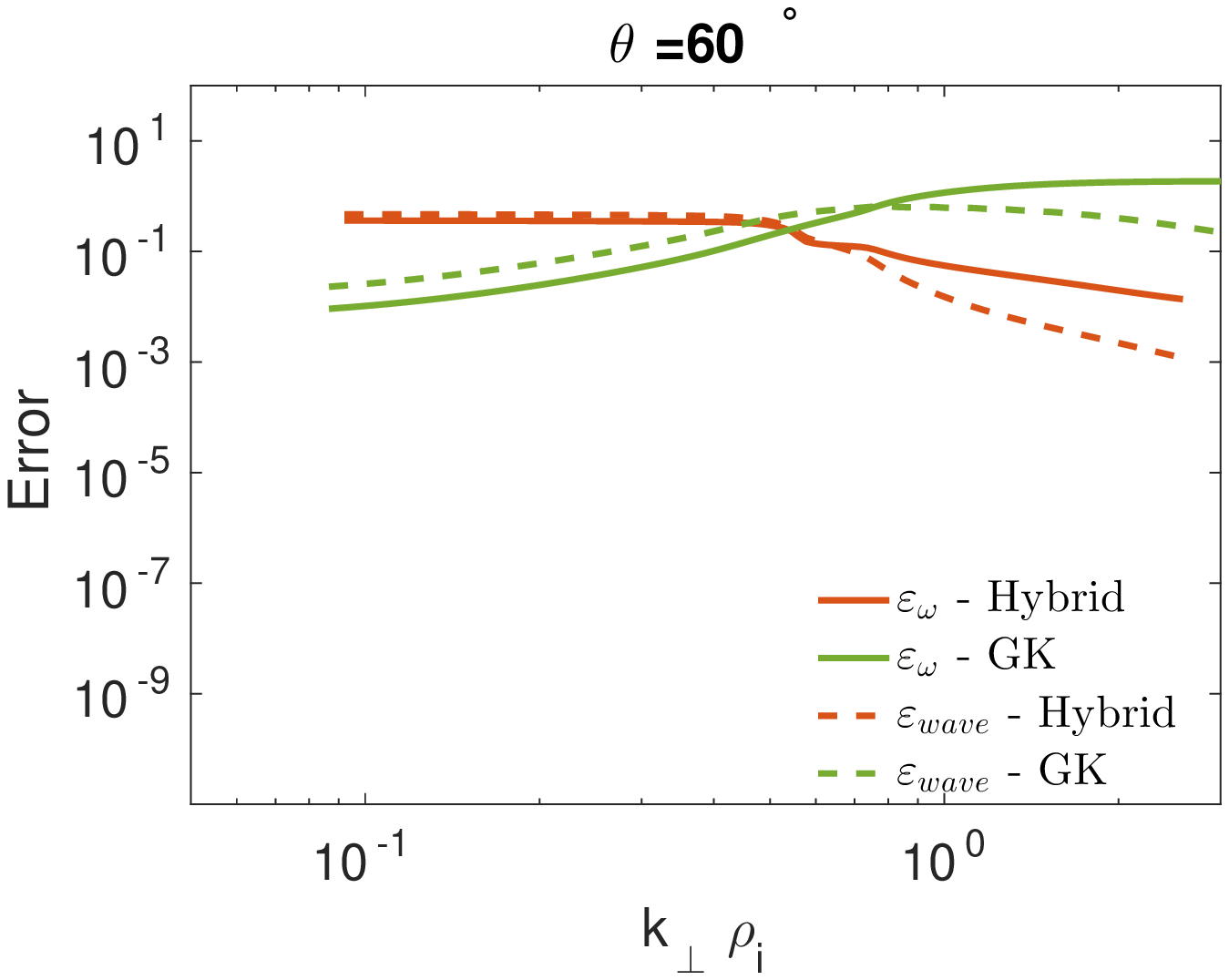}}
  \caption{Errors for Alfv\'{e}n/ion-cyclotron wave with $\theta=60^\circ$, as functions of $k_\perp\rho_i$. Red and green curves are for hybrid and gyrokinetics, respectively. Solid lines denote $\varepsilon_\omega$, and dashed lines denote $\varepsilon_{wave}$.}
\label{fig:error_AIC_60}
\end{figure}

\begin{figure}
  \includegraphics[width=10cm]{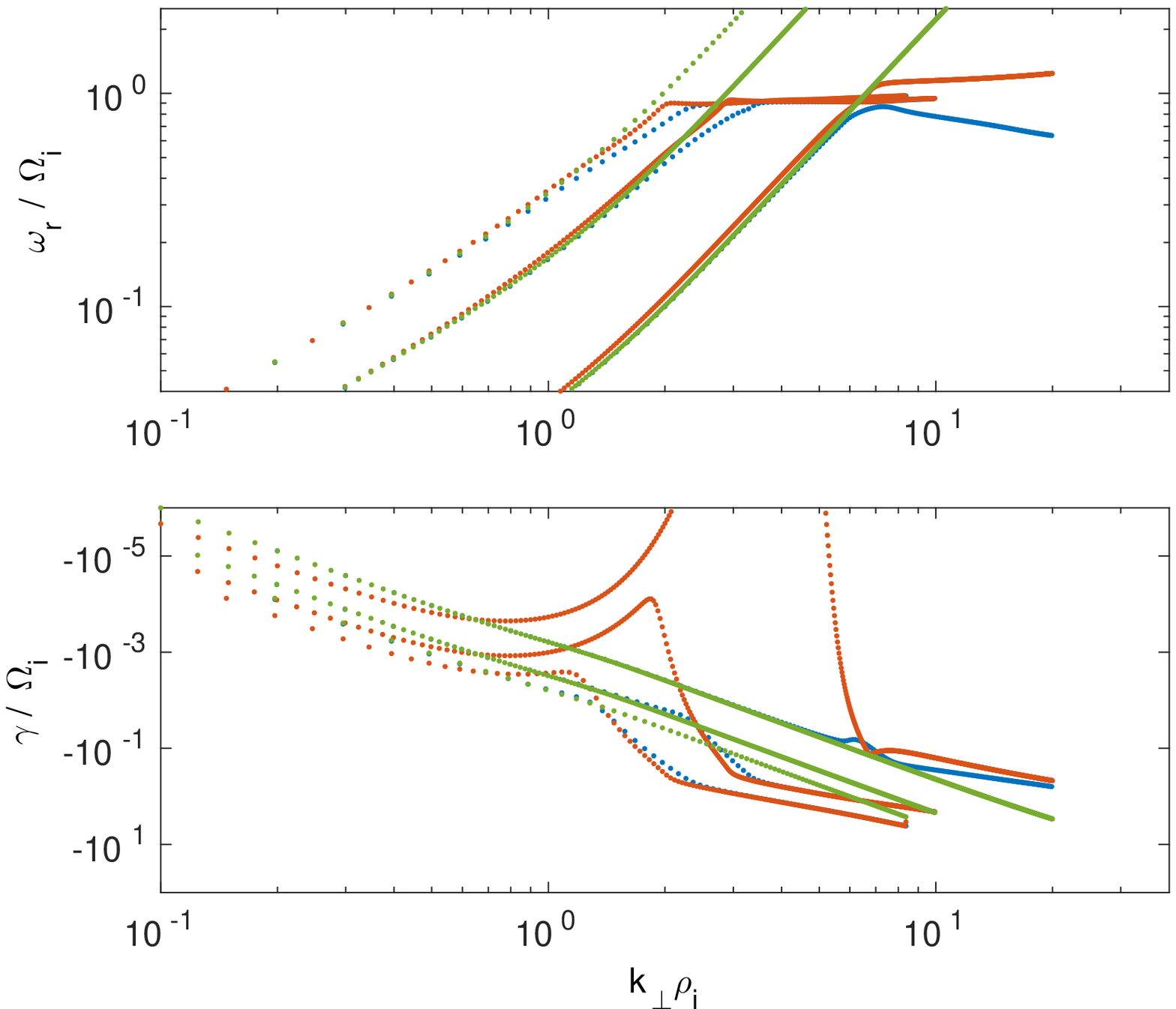}
  \caption{Dispersion relation for Alfv\'{e}n/ion-cyclotron waves for angles of propagation $\theta=80^\circ, 85^\circ, 89^\circ$ (curves from left to right). Blue, red, and green lines are for fully-kinetic, hybrid, and gyrokinetic, respectively. Top panel: real frequency $\omega_r$; bottom panel: damping rate $\gamma$. $k_\perp\rho_i$ is on the horizontal axis. All axis are in logarithmic scale.}
\label{fig:disp_rel_AIC_perp}
\end{figure}

\begin{figure}
  \centerline{\includegraphics[width=10cm]{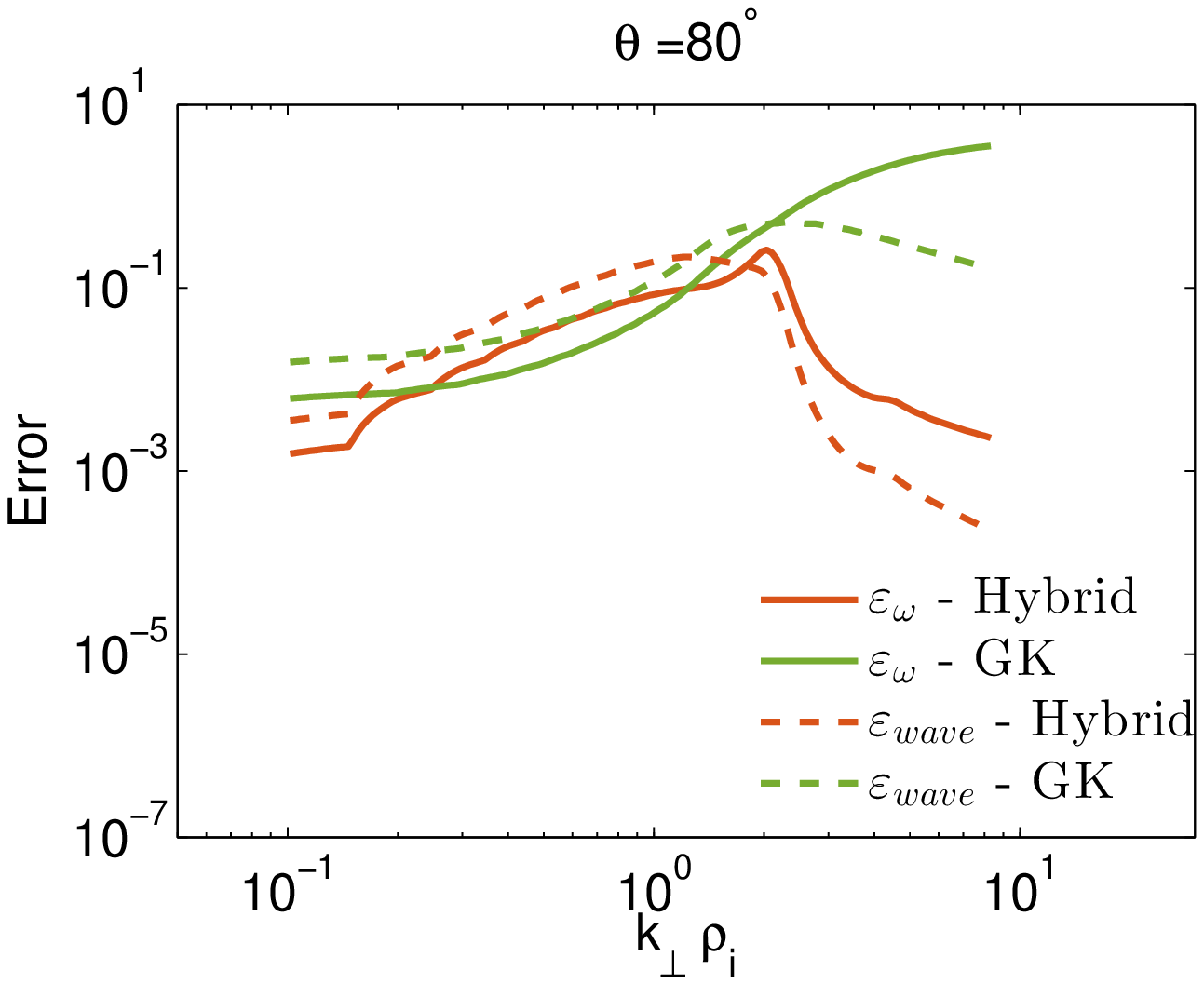}}
  \caption{Errors for Alfv\'{e}n/ion-cyclotron wave with $\theta=80^\circ$, as functions of $k_\perp\rho_i$. Red and green curves are for hybrid and gyrokinetics, respectively. Solid lines denote $\varepsilon_\omega$, and dashed lines denote $\varepsilon_{wave}$.}
\label{fig:error_AIC_80}
\end{figure}

\begin{figure}
  \centerline{\includegraphics[width=10cm]{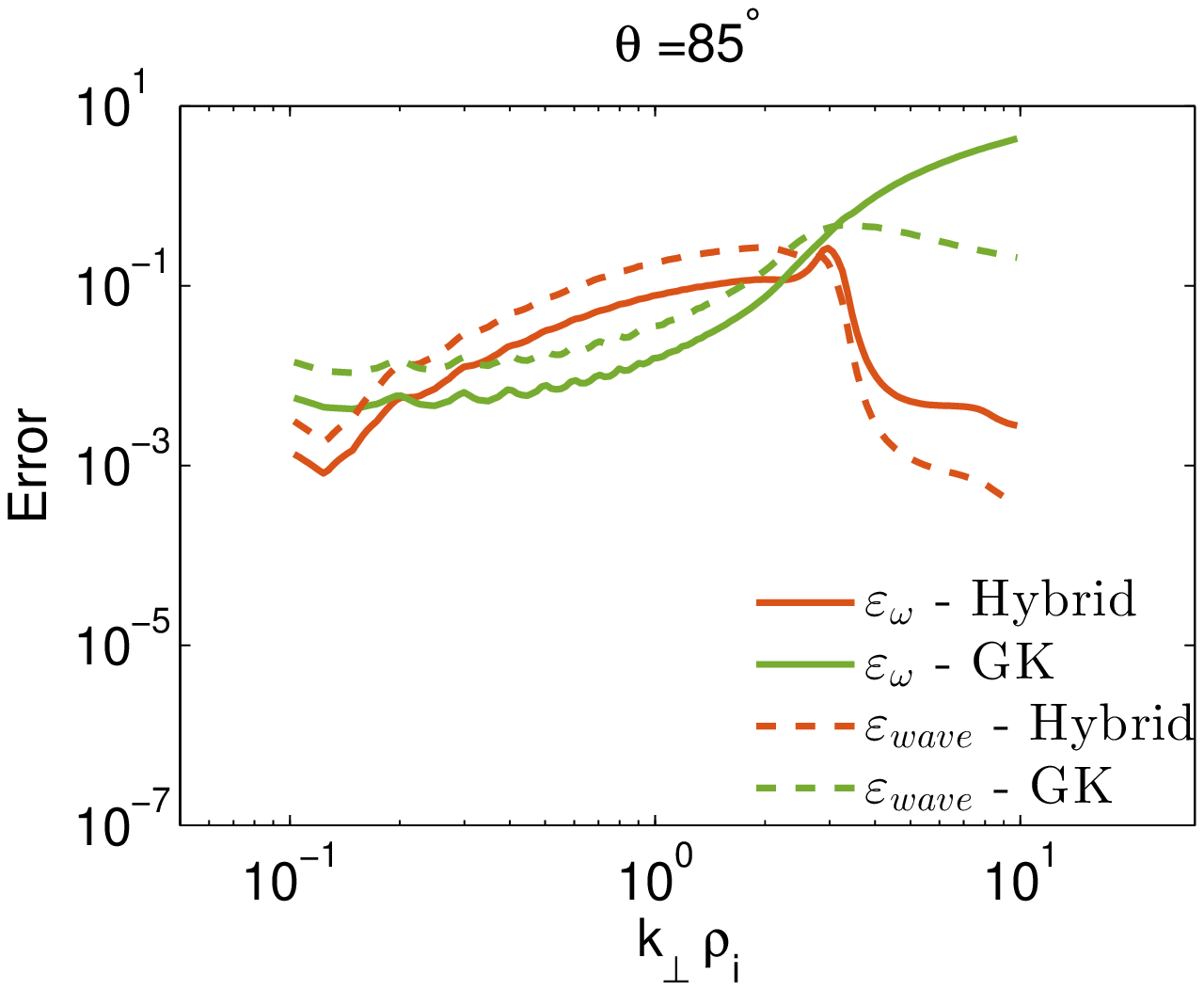}}
  \caption{Errors for Alfv\'{e}n/ion-cyclotron wave with $\theta=85^\circ$, as functions of $k_\perp\rho_i$. Red and green curves are for hybrid and gyrokinetics, respectively. Solid lines denote $\varepsilon_\omega$, and dashed lines denote $\varepsilon_{wave}$.}
\label{fig:error_AIC_85}
\end{figure}

\begin{figure}
  \centerline{\includegraphics[width=10cm]{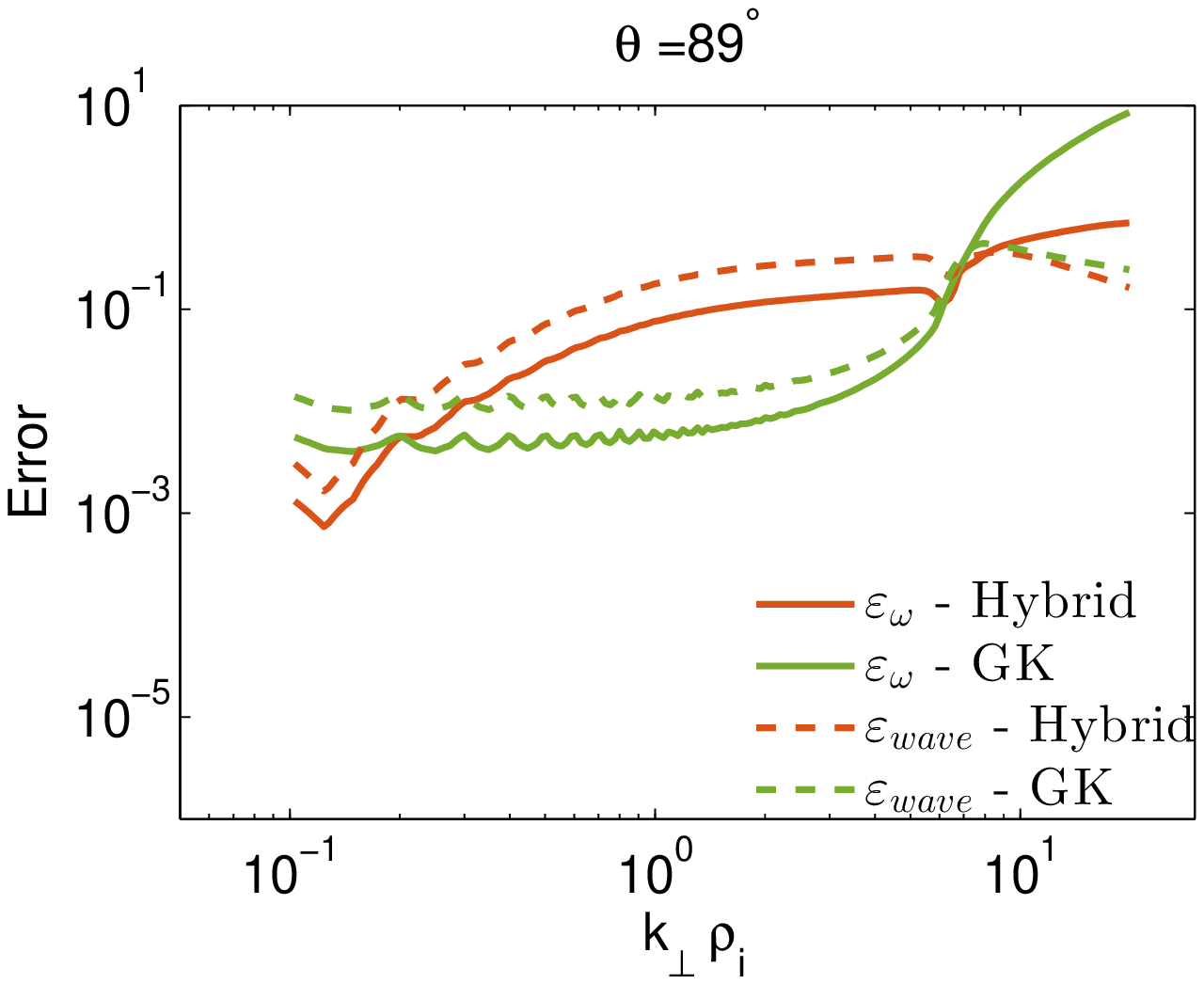}}
  \caption{Errors for Alfv\'{e}n/ion-cyclotron wave with $\theta=89^\circ$, as functions of $k_\perp\rho_i$. Red and green curves are for hybrid and gyrokinetics, respectively. Solid lines denote $\varepsilon_\omega$, and dashed lines denote $\varepsilon_{wave}$.}
\label{fig:error_AIC_89}
\end{figure}

Figure \ref{fig:disp_rel_AIC} shows the dispersion relation for the AIC mode, for angles of propagation $\theta = 10^\circ, 30^\circ, 60^\circ$. Blue, red and green curves are for fully-kinetic, hybrid and gyrokinetic, respectively (the same notation will be used for all following figures). The real frequency $\omega_r$ is shown in the top panel, and the damping rate in the bottom panel. In both panels $k_\perp\rho_i$ is on the horizontal axis (logarithmic scale is used on all axis). The three different propagation angles can be identified from the starting values of $k_\perp\rho_i$. Since all the curves are plotted from $k\rho_i=0.1$, the initial perpendicular wavevector increases with increasing $\theta$. For $\theta=10^\circ$ and $30^\circ$, the hybrid and fully-kinetic solutions are almost indistinguishable. In particular, the saturation of wave frequency is well represented. The corresponding damping rates present a mismatch at low $k$ and agree well at large $k$. A peculiar feature of the $\theta=60^\circ$ solution is that a branch bifurcation happens around $k_\perp\rho_i = 0.6$ for the hybrid solution. The gyrokinetic solution does not capture the correct frequency saturation, for any angle of propagation, as expected. Consequently, the frequencies of the gyrokinetic solutions become much higher than the correct values, even for $k_\perp\rho_i<1$. The gyrokinetic solutions are generally less damped than the correct fully-kinetic solutions (bottom panel).
The errors for the AIC mode solutions are shown in Figures \ref{fig:error_AIC_10}, \ref{fig:error_AIC_30}, \ref{fig:error_AIC_60}, for the cases $\theta=10^\circ$, $30^\circ$, $60^\circ$, respectively. Here, the hybrid model is shown in red, and the gyrokinetic in green.
Solid and dashed lines denote the errors $\varepsilon_\omega$ and $\varepsilon_{wave}$, respectively. One can notice that the errors for the hybrid solutions are lower than the corresponding errors for the gyrokinetic model, for $\theta=10^\circ$ and $30^\circ$. However, which model performs better for $\theta=60^\circ$ depends on $k_\perp\rho_i$
The fact that the gyrokinetic errors in Figure \ref{fig:error_AIC_60} for $\theta=60^\circ$ do not grow very large, and almost saturate for $k_\perp\rho_i>1$, despite the fact that the dispersion relation shows a large mismatch with the fully-kinetic solution, is due to the fact that the mode becomes heavily damped, and therefore the contribution of the real frequency mismatch becomes negligible in the errors.\\
Moving to quasi-perpendicular angles, Figure \ref{fig:disp_rel_AIC_perp} shows the dispersion relation for AIC for $\theta = 80^\circ, 85^\circ, 89^\circ$ (from left to right). Interesting, even at such large propagation angle, the AIC resonate with ions and presents a frequency saturation. Once again the gyrokinetic solution is quite faithful (now both in frequency and damping rate) up to the wavenumbers where the dispersion relation saturates. In this regime, the hybrid solution presents larger errors, in particular close to the elbow of the curves, where the damping rates depart from the Vlasov-Maxwell values (bottom panel). Moreover, in the $\theta = 89^\circ$ case, the saturated value of the hybrid solution is not correct (being larger than $\omega/\Omega_i =1 $). The respective errors are plotted in Figures \ref{fig:error_AIC_80}, \ref{fig:error_AIC_85}, and \ref{fig:error_AIC_89}. It is clear how the gyrokinetic solution gets better and better, with respect to the hybrid, with increasing angle of propagation (only for wavevectors smaller than the value where dispersive effects take place).

\subsection{Ion-acoustic (IA)}
\begin{figure}
  \includegraphics[width=10cm]{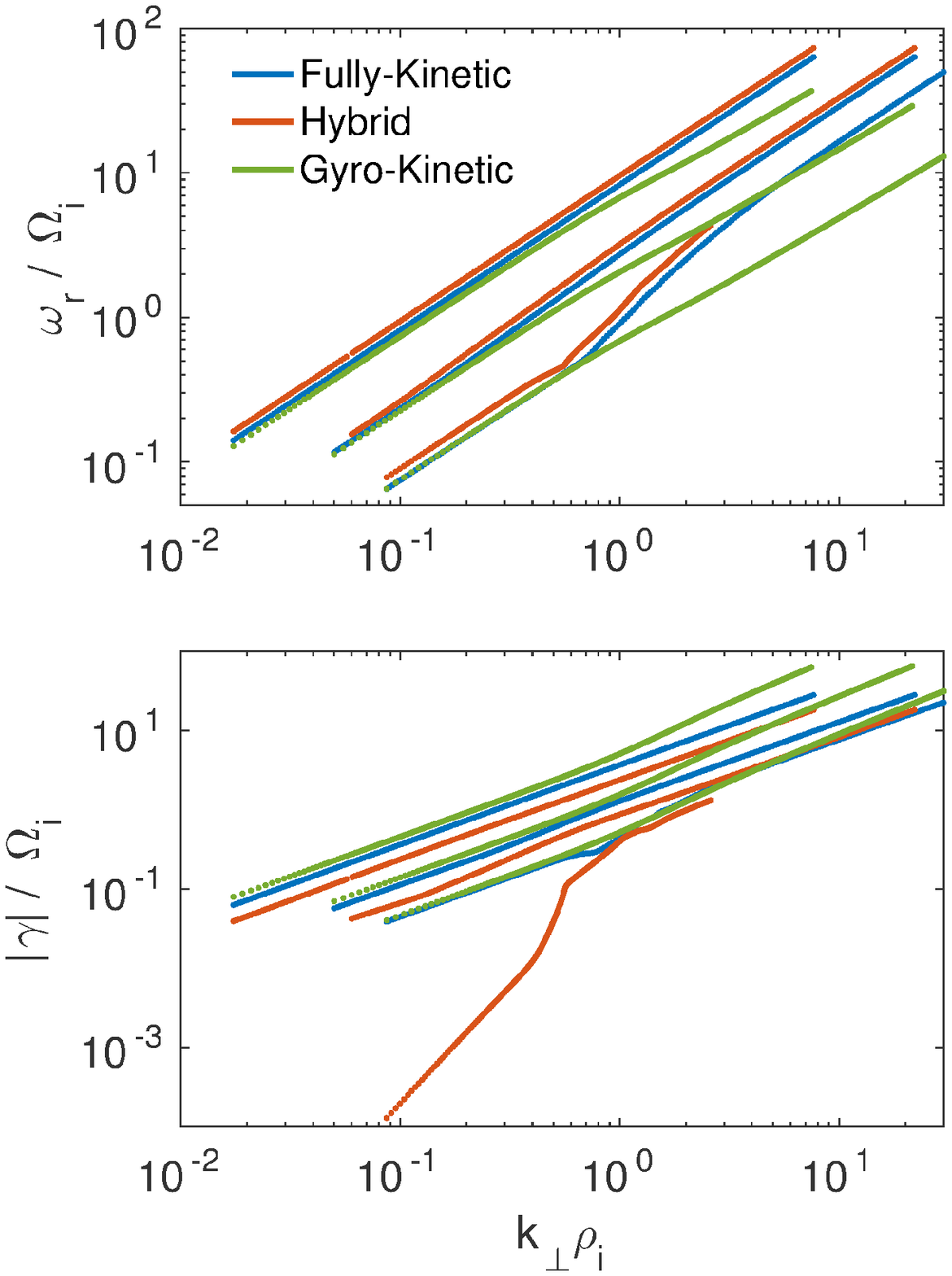}
  \caption{Dispersion relation for ion-acoustic waves for angles of propagation $\theta=10^\circ, 30^\circ, 60^\circ$ (curves from left to right). Blue, red, and green lines are for fully-kinetic, hybrid, and gyrokinetic, respectively. Top panel: real frequency $\omega_r$; bottom panel: damping rate $\gamma$. $k_\perp\rho_i$ is on the horizontal axis. All axis are in logarithmic scale.}
\label{fig:disp_rel_IA}
\end{figure}

\begin{figure}
  \centerline{\includegraphics[width=10cm]{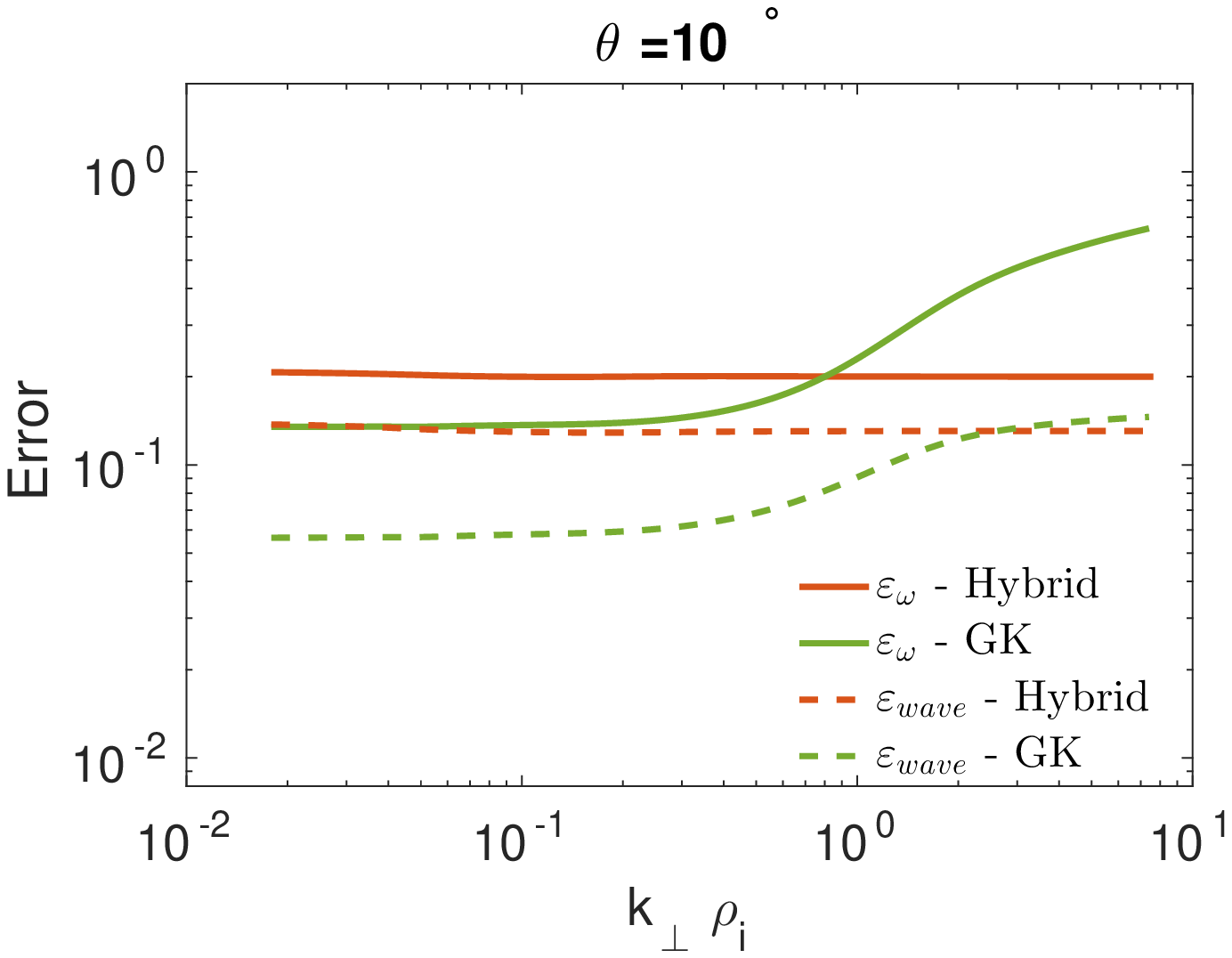}}
  \caption{Errors for ion-acoustic wave with $\theta=10^\circ$, as functions of $k_\perp\rho_i$. Red and green curves are for hybrid and gyrokinetics, respectively. Solid lines denote $\varepsilon_\omega$, and dashed lines denote $\varepsilon_{wave}$.}
\label{fig:error_IA_10}
\end{figure}

\begin{figure}
  \centerline{\includegraphics[width=10cm]{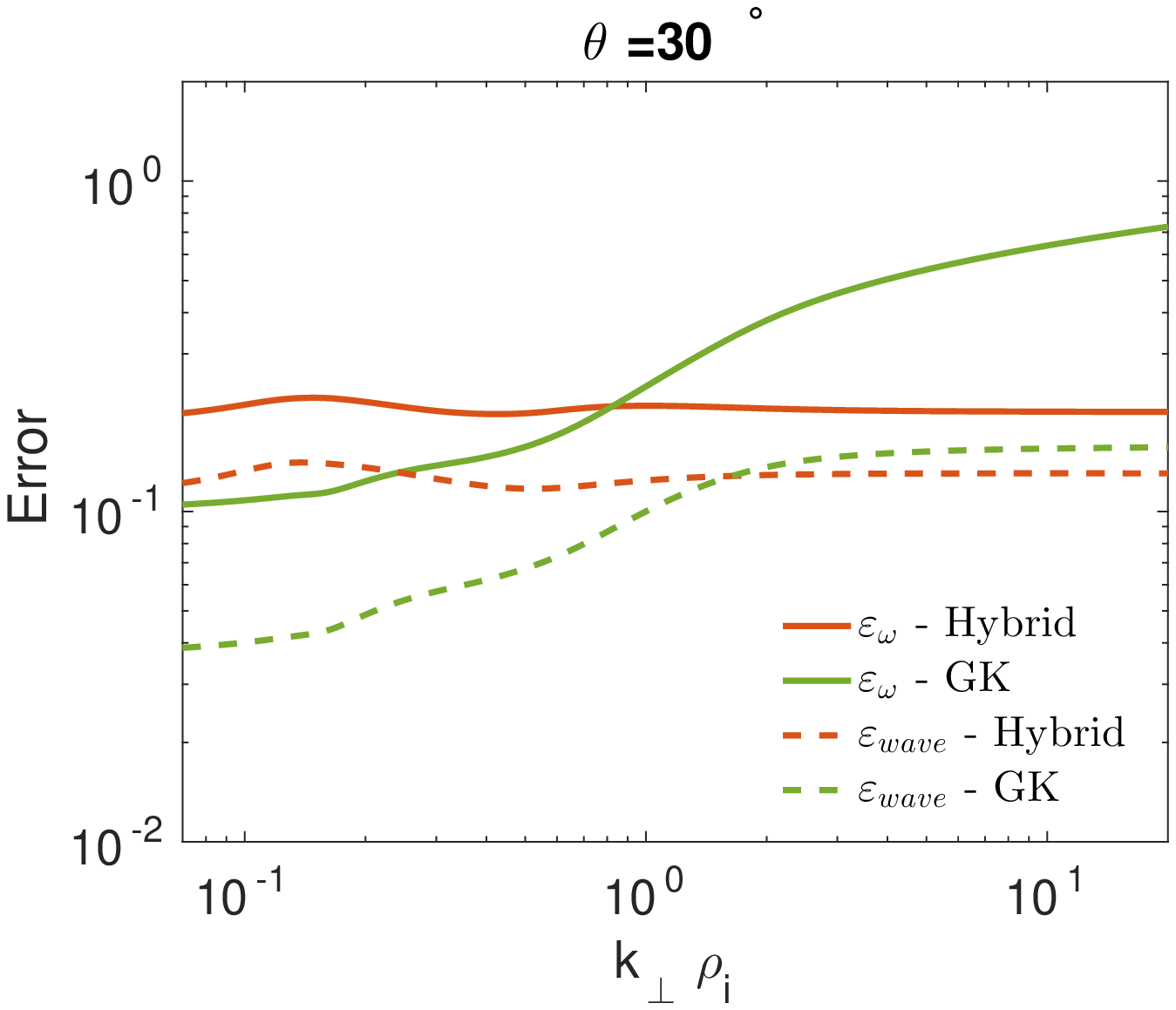}}
  \caption{Errors for ion-acoustic wave with $\theta=30^\circ$, as functions of $k_\perp\rho_i$. Red and green curves are for hybrid and gyrokinetics, respectively. Solid lines denote $\varepsilon_\omega$, and dashed lines denote $\varepsilon_{wave}$.}
\label{fig:error_IA_30}
\end{figure}

\begin{figure}
  \centerline{\includegraphics[width=10cm]{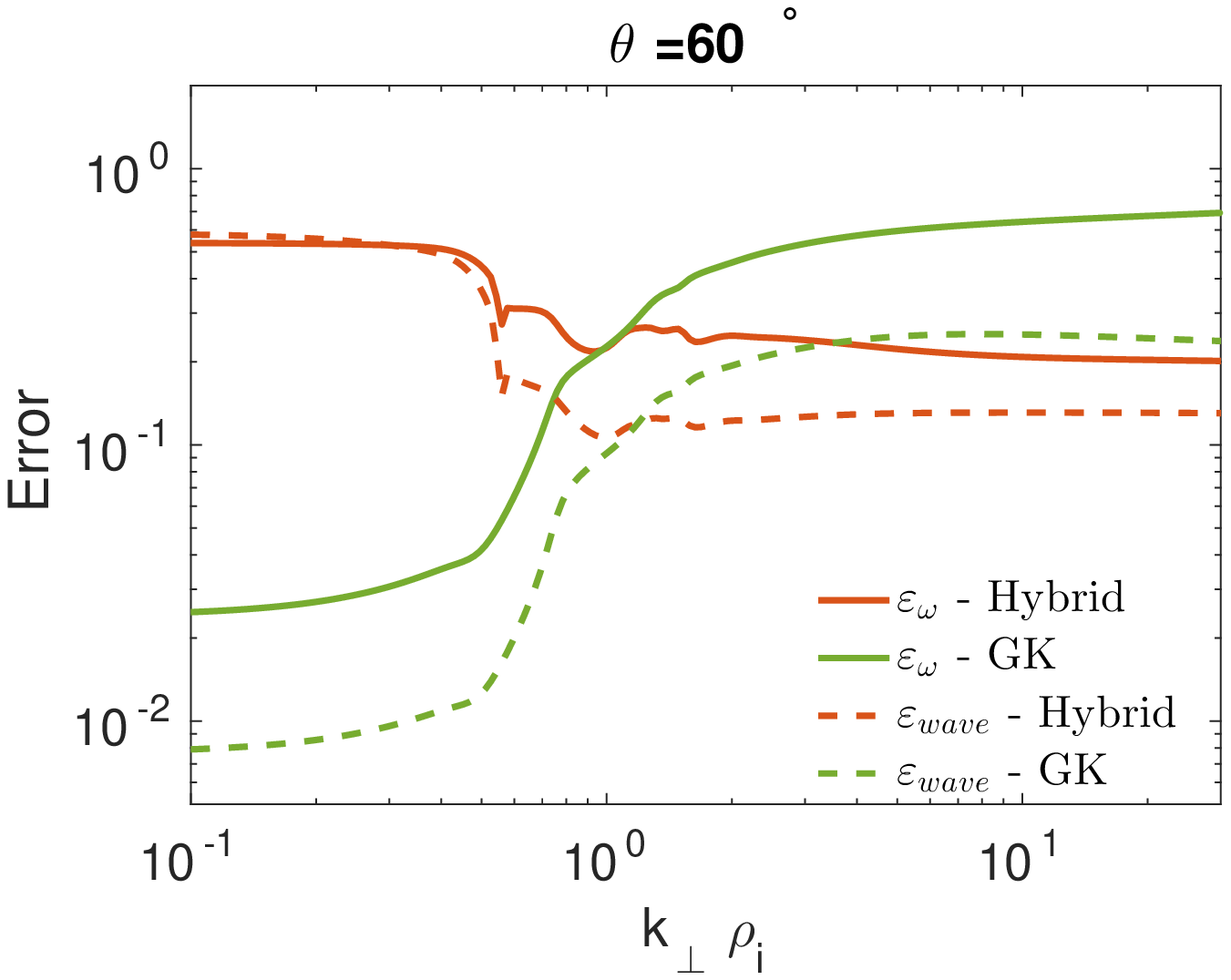}}
  \caption{Errors for ion-acoustic wave with $\theta=60^\circ$, as functions of $k_\perp\rho_i$. Red and green curves are for hybrid and gyrokinetics, respectively. Solid lines denote $\varepsilon_\omega$, and dashed lines denote $\varepsilon_{wave}$.}
\label{fig:error_IA_60}
\end{figure}

\begin{figure}
  \includegraphics[width=10cm]{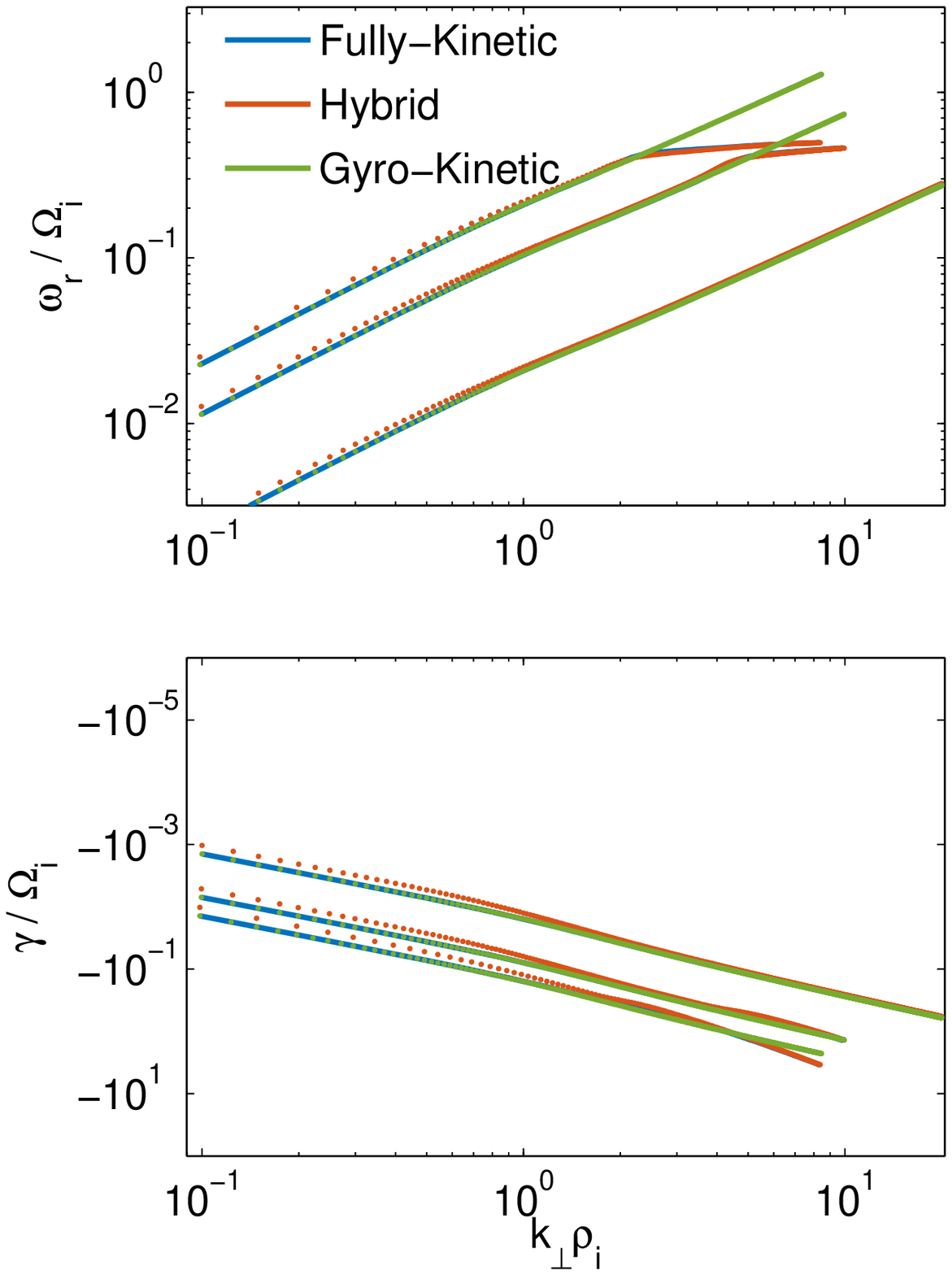}
  \caption{Dispersion relation for ion-acoustic waves for angles of propagation $\theta=80^\circ, 85^\circ, 89^\circ$ (curves from left to right). Blue, red, and green lines are for fully-kinetic, hybrid, and gyrokinetic, respectively. Top panel: real frequency $\omega_r$; bottom panel: damping rate $\gamma$. $k_\perp\rho_i$ is on the horizontal axis. All axis are in logarithmic scale.}
\label{fig:disp_rel_IA_perp}
\end{figure}

\begin{figure}
  \centerline{\includegraphics[width=10cm]{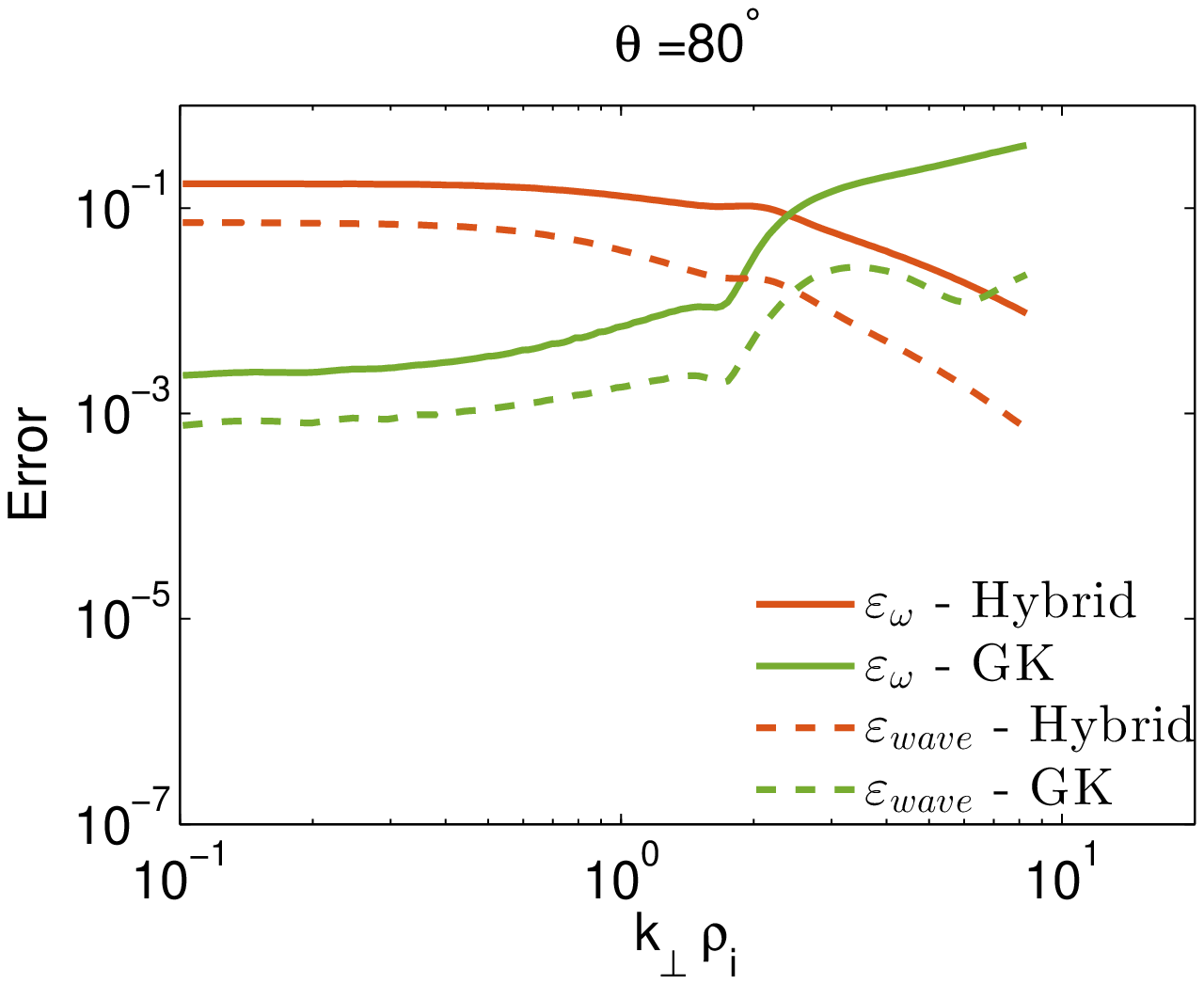}}
  \caption{Errors for ion-acoustic wave with $\theta=80^\circ$, as functions of $k_\perp\rho_i$. Red and green curves are for hybrid and gyrokinetics, respectively. Solid lines denote $\varepsilon_\omega$, and dashed lines denote $\varepsilon_{wave}$.}
\label{fig:error_IA_80}
\end{figure}

\begin{figure}
  \centerline{\includegraphics[width=10cm]{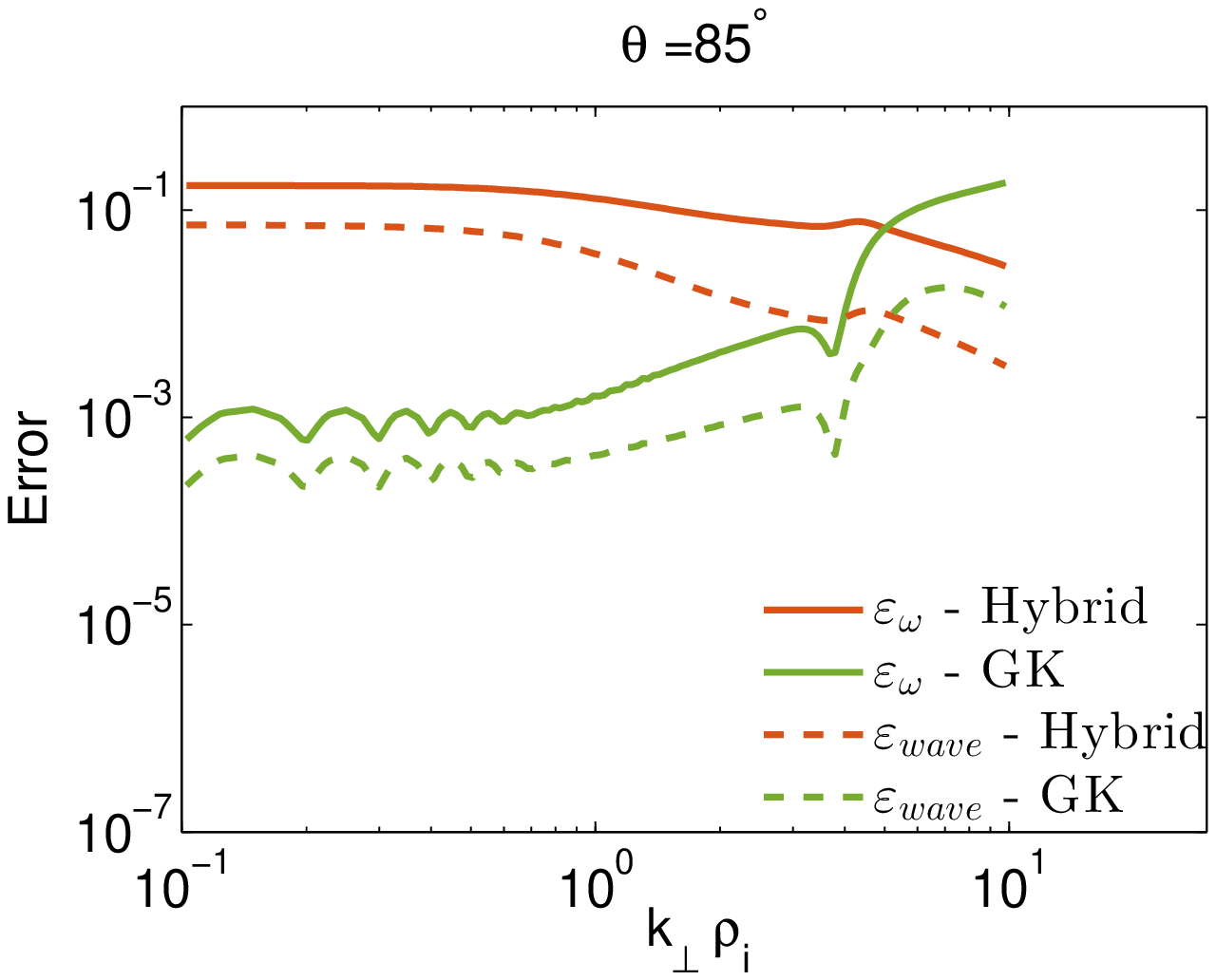}}
  \caption{Errors for ion-acoustic wave with $\theta=85^\circ$, as functions of $k_\perp\rho_i$. Red and green curves are for hybrid and gyrokinetics, respectively. Solid lines denote $\varepsilon_\omega$, and dashed lines denote $\varepsilon_{wave}$.}
\label{fig:error_IA_85}
\end{figure}

\begin{figure}
  \centerline{\includegraphics[width=10cm]{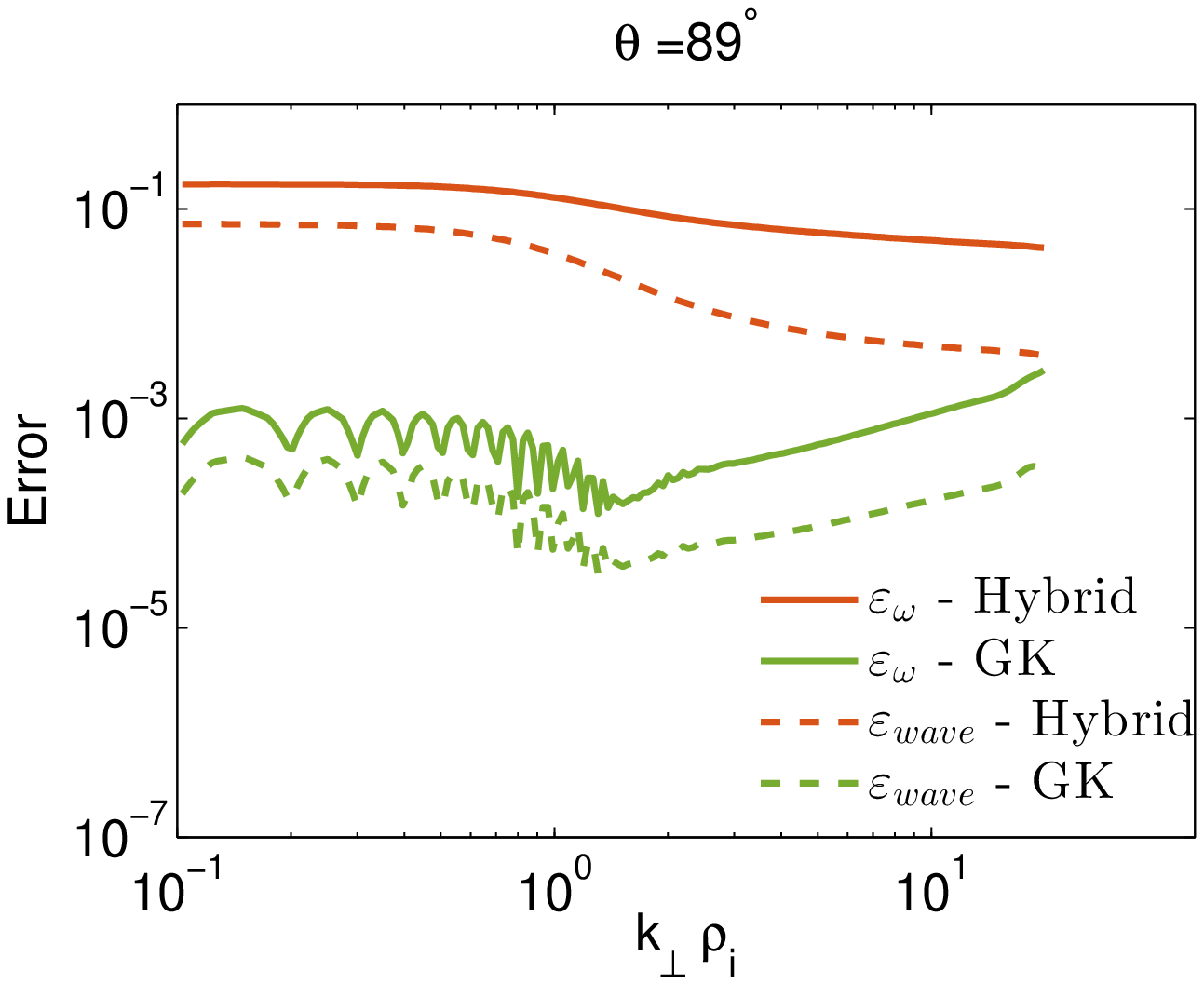}}
  \caption{Errors for ion-acoustic wave with $\theta=89^\circ$, as functions of $k_\perp\rho_i$. Red and green curves are for hybrid and gyrokinetics, respectively. Solid lines denote $\varepsilon_\omega$, and dashed lines denote $\varepsilon_{wave}$.}
\label{fig:error_IA_89}
\end{figure}

The dispersion relation for the IA mode is shown in Figure \ref{fig:disp_rel_IA}, with the same format as in the previous Section. One can distinguish the following features. Regarding the real frequency (top panel), there is again a good agreement between hybrid and fully-kinetic; the gyro-kinetic frequency also presents a similar agreement, but only for small $k$ values. Damping rates present larger mismatches. Notably, the dispersion relation for the case $\theta=60^\circ$ presents an 'elbow' for $k_\perp\rho_i\sim 1$ that is captured by the hybrid model.
We show the errors $\varepsilon_\omega$ and  $\varepsilon_{wave}$ for $\theta=10^\circ$, $\theta=30^\circ$, $\theta=60^\circ$ respectively in Figures \ref{fig:error_IA_10}, \ref{fig:error_IA_30}, and \ref{fig:error_IA_60} (same format as previous Section). 
The errors are now generally smaller for gyrokinetic at small $k_\perp\rho_i$, and increasing with wavector, and almost constant in value fo rhte hybrid solutions. 
The dispersion relation plots for quasi-perpendicular angles are shown in Figure \ref{fig:disp_rel_IA_perp}. Similarly to the AIC case, the gyrokinetic model reproduces more faithfully the Vlasov-Maxwell solutions, until dispersive effects take place. A small but noticeable mismatch in the sound speed can be noted for the hybrid solution, that is shifted vertically. The only curve where the gyrokinetic and the Vlasov-Maxwell solution agree for the whole range of wavevectors considered is for $\theta=89^\circ$.
The corresponding errors are shown in Figures \ref{fig:error_IA_80}, \ref{fig:error_IA_85}, and \ref{fig:error_IA_89}. It is interesting to notice how the relative accuracy of the two models strongly depend on the wavevector. For instance, at $\theta = 80^\circ$ the gyrokinetic solution is about two order of magnitudes more accurate than the hybrid for $k_\perp\rho_i = 0.1$, but starts to become less accurate for
$k_\perp\rho_i >2 $. Also, it is interesting that the hybrid solution errors decrease with increasing wavevector. That is, the model is more accurate in the dispersive regions of the curves.

\subsection{Fast mode}
\begin{figure}
  \includegraphics[width=10cm]{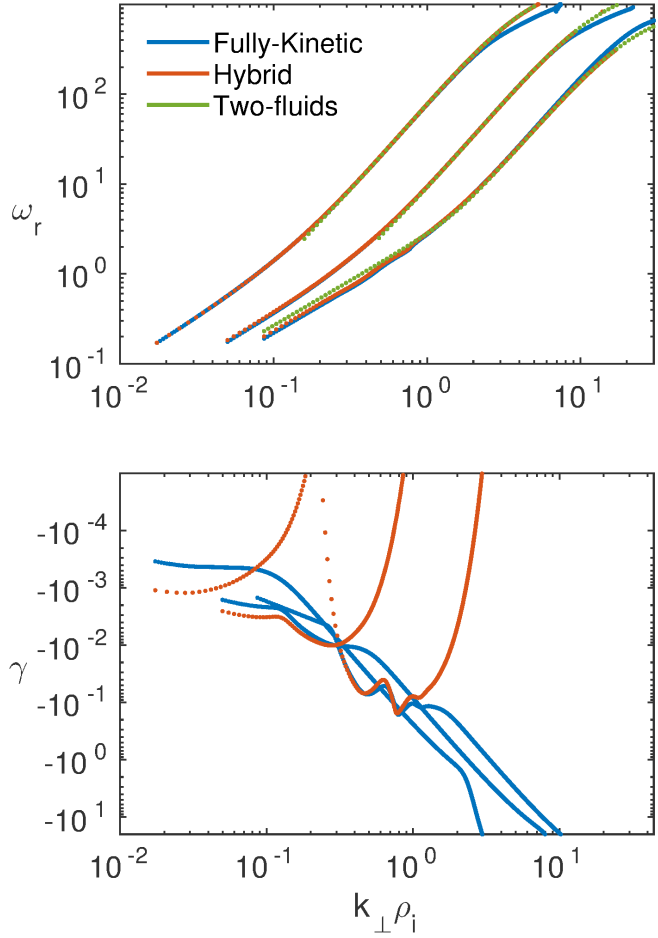}
  \caption{Dispersion relation for fast waves for angles of propagation $\theta=10^\circ, 30^\circ, 60^\circ$ (curves from left to right). Blue, red, and green lines are for fully-kinetic, hybrid, and two-fluid, respectively. Top panel: real frequency $\omega_r$ in logarithmic scale; bottom panel: damping rate $\gamma$ in logarithmic scale. $k_\perp\rho_i$ is on the horizontal axis in logarithmic scale.}
\label{fig:disp_rel_FW}
\end{figure}

It is well-known that the fast mode solution is factored out from the gyrokinetic model. In this sense, a comparison of gyrokinetics with the other the two models is meaningless. 
In Figure \ref{fig:disp_rel_FW} we compare the dispersion relations of the fully-kinetic, the hybrid, and the two-fluids models for $\theta=10^\circ$, $\theta=30^\circ$, $\theta=60^\circ$. The dispersion relation for the latter follows the derivation of \citet{zhao14}. The three solutions are in a very good agreement for the real frequency (top panel), where a slight deviation with respect to the correct solution occurs only at large $k_\perp$. Interestingly, the two-fluids solution is a very good proxy for both hybrid and full-kinetic solutions, in this regime.
The bottom panel shows the corresponding damping rate (the two-fluids model is not shown since its solutions are always real). Once again, as expected, the hybrid model underestimates the damping rates, although there is a certain regime, approximately for $k\rho_i\leq 1$, where the red and blue curves have a good agreement.

\subsection{Quasi-perpendicular propagation}
\begin{figure}
  \includegraphics[width=10cm]{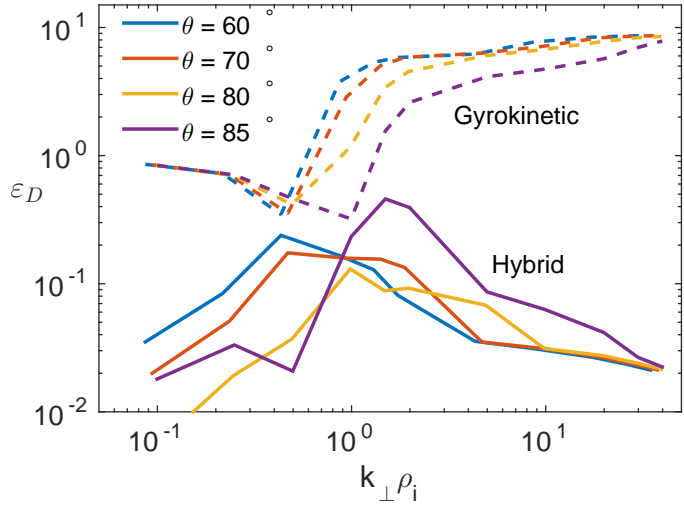}
  \caption{Error $\varepsilon_D$, as a function of $k_\perp\rho_i$, for angles of propagation $\theta=60^\circ, 70^\circ, 80^\circ, 85^\circ$ (blue, red, yellow, and magenta). Dashed and solid lines are for gyrokinetics and hybrid, respectively.}
\label{fig:error_D}
\end{figure}

\begin{figure}
  \includegraphics[width=15cm]{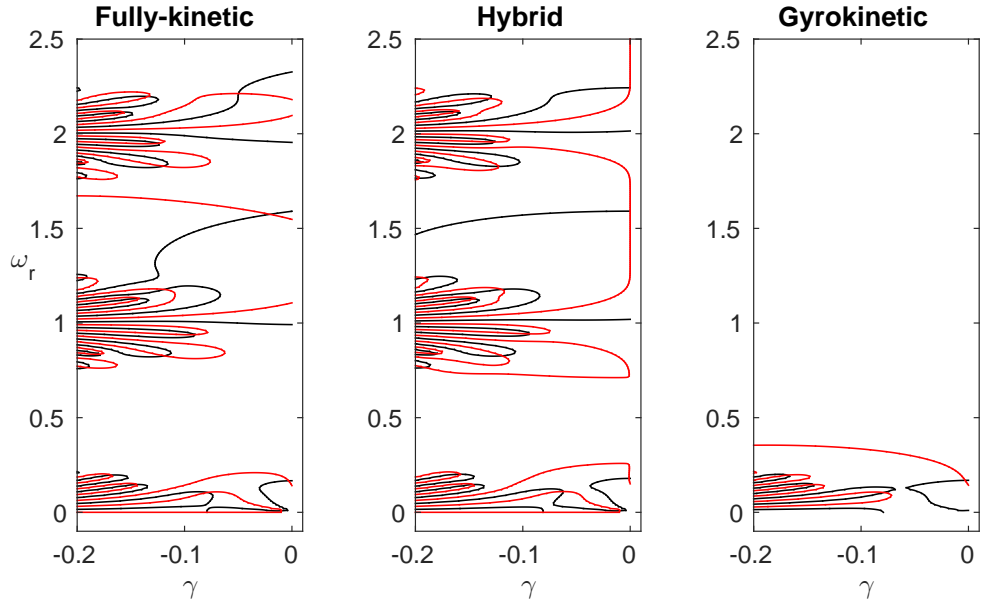}
  \caption{Real (black) and imaginary (red) part of $f(\omega)=\det(\bf{D})$ for $\theta=85^\circ$, $k\rho_i=1$. Left panel: fully-kinetic; middle: hybrid; right: gyrokinetic. The intersection between black and red curves identify normal modes.}
\label{fig:modes_contours}
\end{figure}

In this section we perform the analysis of the error $\varepsilon_D$ of Eq. (\ref{eps_D}). As we discussed above, in this regime the goal is not to compare the solutions for normal modes, but rather to provide a single measure of the error that captures the overall distance between the models in term of their pseudospectrum. This approach is justified by the non-normality of the linear operators, and by the fact that a multitude of modes appear in this regime, and it is not obvious why any of those should play a particular role in the plasma dynamics. 
Figure \ref{fig:error_D} shows the error $\varepsilon_D$ for angles of propagation $\theta=60^\circ$, $\theta=70^\circ$, $\theta=80^\circ$, and $\theta=85^\circ$ (respectively in blue, red, yellow, and magenta). The solid lines are for the hybrid model, and the dashed lines for the gyrokinetic.
A very interesting trend can be noticed. The errors for the gyrokinetic model have a sharp increase, of almost an order of magnitude, around $k_\perp\rho_i=1$, while the hybrid model error tend to increase for $k_\perp\rho_i<1$ and to decrease for $k_\perp\rho_i>1$. Overall, the hybrid model performs much better than the gyrokinetic. This is not surprising because, as it was already evident by inspection of Figures \ref{fig:pseudo} and \ref{fig:pseudo_k_1_85}, the gyrokinetic model simply misses many roots of the fully-kinetic dispersion relation. It is important to realize that the modes that are not captured by this model are not 'heavily-damped' with respect to the modes that are captured (but they have higher frequencies).
One could argue that the error $\varepsilon_D$ is defined and constructed in a rather arbitrary way. Although this is certainly true, we think that it is still very informative on how much a reduced model is able to reproduce the correct linear evolution of the system. as we mentioned, the ability of gyrokinetics to describe solar wind turbulence in the regime $k_\perp\gg k_{||}$ has been stressed out in many recent papers \citep{schekochihin09, tenbarge13}.
Although this might certainly be true, we believe that it is informative to know what is left out from the model, in terms of normal modes. In Figure \ref{fig:modes_contours} we compare the three models (fully-kinetic to the left, hybrid in the middle, gyrokinetic to the right) for the case $k\rho_i=1$, and $\theta=85^\circ$. Here, we do not show the normal modes, but the curves $\Re(f)=0$ (black), and $\Im(f)=0$ (red) in the domain $-0.2<\gamma<0$ and $0<\omega_r<2.5$. The normal modes can be inferred as the points where black and red curves intersect. This figure reinforces what is well-known about the models: gyrokinetics factors out high-frequency modes (irrespective of their damping rate), and the hybrid model does not properly damp some of the modes (the ones that undergo electron Landau damping). 

\section{Conclusions}\label{sec:conclusions}
Reduced kinetic models are extensively used in plasma physics, in lieu of the more complete Vlasov-Maxwell system. In this paper, we have compared the hybrid (ion-kinetic and electron-fluid), the gyrokinetic, and the fully-kinetic models, for a wide range of wavevectors and propagation angles, in linear theory.
We have chosen one test case with parameters relevant to the study of solar wind turbulence. We have compared both dispersion relations of single normal modes, for angles of propagation $\theta = 10^\circ, 30^\circ, 60^\circ, 80^\circ, 85^\circ, 89^\circ$ and pseudospectra, for   $\theta = 60^\circ, 70^\circ, 80^\circ, 85^\circ$. The latter method is justified by the multitude of modes that appear in the quasi-perpendicular case, which makes it non-trivial to identify which one plays a predominant role. It is important to remember that that the gyrokinetic model has been employed also outside its nominal range of validity ($k_{||}\ll k_\perp$).
We note that the Vlasov-Maxwell linear theory has recently been investigated by \citet{sahraoui12}, for very oblique propagation angles. 
Our findings can be summarized as follows. The hybrid solutions present a reasonable agreement with the Vlasov-Maxwell system across all propagation angles, the largest disagreement being for $\theta=89^\circ$. Interestingly, the errors decrease in the region when the waves resonate with ions, and the dispersion curves saturate. One major mismatch of the hybrid model happens for $\theta = 60^\circ$, where a clear matching of the hybrid AIC and IA modes with the corresponding Vlasov-Maxwell modes is not possible due to branch switching.  
As expected the gyrokinetic model presents larger errors for angles of propagation less than $80^\circ$, particularly for $k_\perp\rho_i>1$ However, also for quasi-perpendicular propagation, the errors increase abruptly as soon as the waves become resonant with ions.
In conclusion, it appears that the extension of the gyrokinetic model to the electron regime ($k_\perp\rho_i>1$) is not justified, for angles of propagation smaller than about $85^\circ$. This is also apparent from looking at the analysis of the error based on the pseudospectra. Indeed, it appears that the hybrid model should be preferred, for its ability to represent most of the normal modes (certainly in the ion regime $k_\perp\rho_i<1$ and, to a certain extent, in the electron regime as well). However, we notice that the spurious low value of damping rates of some hybrid solutions might result in numerical instabilities, due to the accumulation of energy that is not properly damped. In this respect, a promising numerical approach consists in devising models that factor out Langmuir oscillations, still retaining the fully-kinetic nature of the solution (see, e.g. \citep{degond12, chen14, tronci15}).

\newpage

\clearpage

\clearpage

\clearpage

%\appendix
%\section{}\label{appA}

\bibliographystyle{jpp}

\newpage
\clearpage
%\bibliography{camporeale_bib}

\begin{thebibliography}{46}
\expandafter\ifx\csname natexlab\endcsname\relax\def\natexlab#1{#1}\fi

\bibitem[Boldyrev {\em et~al.\/}(2013)Boldyrev, Horaites, Xia \&
  Perez]{boldyrev13}
{\sc Boldyrev, Stanislav, Horaites, Konstantinos, Xia, Qian \& Perez,
  Jean~Carlos} 2013 Toward a theory of astrophysical plasma turbulence at
  subproton scales. {\em The Astrophysical Journal\/} {\bf 777}~(1), 41.

\bibitem[Brizard \& Hahm(2007)]{brizard07}
{\sc Brizard, AJa \& Hahm, TSb} 2007 Foundations of nonlinear gyrokinetic
  theory. {\em Reviews of modern physics\/} {\bf 79}~(2), 421.

\bibitem[Camporeale(2012)]{camporeale12}
{\sc Camporeale, Enrico} 2012 Nonmodal linear theory for space plasmas. {\em
  Space science reviews\/} {\bf 172}~(1-4), 397--409.

\bibitem[Camporeale \& Burgess(2011)]{camporeale11}
{\sc Camporeale, Enrico \& Burgess, David} 2011 The dissipation of solar wind
  turbulent fluctuations at electron scales. {\em The Astrophysical Journal\/}
  {\bf 730}~(2), 114.

\bibitem[Camporeale {\em et~al.\/}(2009)Camporeale, Burgess \&
  Passot]{camporeale09}
{\sc Camporeale, Enrico, Burgess, David \& Passot, Thierry} 2009 Transient
  growth in stable collisionless plasma. {\em Physics of Plasmas
  (1994-present)\/} {\bf 16}~(3), 030703.

\bibitem[Camporeale {\em et~al.\/}(2016)Camporeale, Delzanno, Bergen \&
  Moulton]{camporeale16}
{\sc Camporeale, E, Delzanno, GL, Bergen, BK \& Moulton, JD} 2016 On the
  velocity space discretization for the vlasov--poisson system: Comparison
  between implicit hermite spectral and particle-in-cell methods. {\em Computer
  Physics Communications\/} {\bf 198}, 47--58.

\bibitem[Camporeale {\em et~al.\/}(2010)Camporeale, Passot \&
  Burgess]{camporeale10}
{\sc Camporeale, Enrico, Passot, Thierry \& Burgess, David} 2010 Implications
  of a non-modal linear theory for the marginal stability state and the
  dissipation of fluctuations in the solar wind. {\em The Astrophysical
  Journal\/} {\bf 715}~(1), 260.

\bibitem[Chen \& Chacon(2014)]{chen14}
{\sc Chen, Guangye \& Chacon, Luis} 2014 An energy-and charge-conserving,
  nonlinearly implicit, electromagnetic 1d-3v vlasov--darwin particle-in-cell
  algorithm. {\em Computer Physics Communications\/} {\bf 185}~(10),
  2391--2402.

\bibitem[Cheng \& Johnson(1999)]{cheng99}
{\sc Cheng, CZ \& Johnson, Jay~R} 1999 A kinetic-fluid model. {\em Journal of
  Geophysical Research: Space Physics\/} {\bf 104}~(A1), 413--427.

\bibitem[Degond {\em et~al.\/}(2012)Degond, Deluzet \& Savelief]{degond12}
{\sc Degond, Pierre, Deluzet, Fabrice \& Savelief, Dominique} 2012 Numerical
  approximation of the euler--maxwell model in the quasineutral limit. {\em
  Journal of Computational Physics\/} {\bf 231}~(4), 1917--1946.

\bibitem[Dendy(1995)]{dendy95}
{\sc Dendy, Richard~O} 1995 {\em Plasma physics: an introductory course\/}.
  Cambridge University Press.

\bibitem[Franci {\em et~al.\/}(2015)Franci, Verdini, Matteini, Landi \&
  Hellinger]{franci15a}
{\sc Franci, Luca, Verdini, Andrea, Matteini, Lorenzo, Landi, Simone \&
  Hellinger, Petr} 2015 Solar wind turbulence from mhd to sub-ion scales:
  High-resolution hybrid simulations. {\em The Astrophysical Journal Letters\/}
  {\bf 804}~(2), L39.

\bibitem[Friedman \& Carter(2014)]{friedman14}
{\sc Friedman, Brett \& Carter, Troy~A} 2014 Linear technique to understand
  non-normal turbulence applied to a magnetized plasma. {\em Physical review
  letters\/} {\bf 113}~(2), 025003.

\bibitem[Gary(2005)]{gary_book}
{\sc Gary, S~Peter} 2005 {\em Theory of space plasma microinstabilities\/}.
  Cambridge university press.

\bibitem[Goswami {\em et~al.\/}(2005)Goswami, Passot \& Sulem]{goswami05}
{\sc Goswami, Priyanka, Passot, T \& Sulem, PL} 2005 A landau fluid model for
  warm collisionless plasmas. {\em Physics of Plasmas (1994-present)\/} {\bf
  12}~(10), 102109.

\bibitem[Haynes {\em et~al.\/}(2014)Haynes, Burgess \& Camporeale]{haynes14}
{\sc Haynes, Christopher~T, Burgess, David \& Camporeale, Enrico} 2014
  Reconnection and electron temperature anisotropy in sub-proton scale plasma
  turbulence. {\em The Astrophysical Journal\/} {\bf 783}~(1), 38.

\bibitem[Haynes {\em et~al.\/}(2015)Haynes, Burgess, Camporeale \&
  Sundberg]{haynes15}
{\sc Haynes, Christopher~T, Burgess, David, Camporeale, Enrico \& Sundberg,
  Torbjorn} 2015 Electron vortex magnetic holes: A nonlinear coherent plasma
  structure. {\em Physics of Plasmas (1994-present)\/} {\bf 22}~(1), 012309.

\bibitem[Howes {\em et~al.\/}(2006)Howes, Cowley, Dorland, Hammett, Quataert \&
  Schekochihin]{howes06}
{\sc Howes, Gregory~G, Cowley, Steven~C, Dorland, William, Hammett, Gregory~W,
  Quataert, Eliot \& Schekochihin, Alexander~A} 2006 Astrophysical
  gyrokinetics: Basic equations and linear theory. {\em The Astrophysical
  Journal\/} {\bf 651}~(1), 590.

\bibitem[Howes {\em et~al.\/}(2011)Howes, TenBarge, Dorland, Quataert,
  Schekochihin, Numata \& Tatsuno]{howes11}
{\sc Howes, Gregory~G, TenBarge, Jason~M, Dorland, William, Quataert, Eliot,
  Schekochihin, Alexander~A, Numata, Ryusuke \& Tatsuno, Tomoya} 2011
  Gyrokinetic simulations of solar wind turbulence from ion to electron scales.
  {\em Physical review letters\/} {\bf 107}~(3), 035004.

\bibitem[Hunana {\em et~al.\/}(2013)Hunana, Goldstein, Passot, Sulem, Laveder
  \& Zank]{hunana13}
{\sc Hunana, Peter, Goldstein, ML, Passot, T, Sulem, PL, Laveder, D \& Zank,
  GP} 2013 Polarization and compressibility of oblique kinetic alfv{\'e}n
  waves. {\em The Astrophysical Journal\/} {\bf 766}~(2), 93.

\bibitem[{Kiyani} {\em et~al.\/}(2013){Kiyani}, {Chapman}, {Sahraoui}, {Hnat},
  {Fauvarque} \& {Khotyaintsev}]{kiyani13}
{\sc {Kiyani}, K.~H., {Chapman}, S.~C., {Sahraoui}, F., {Hnat}, B.,
  {Fauvarque}, O. \& {Khotyaintsev}, Y.~V.} 2013 {Enhanced Magnetic
  Compressibility and Isotropic Scale Invariance at Sub-ion Larmor Scales in
  Solar Wind Turbulence}. {\em The Astrophysical Journal\/} {\bf 763}, 10.

\bibitem[{Lion} {\em et~al.\/}(2016){Lion}, {Alexandrova} \&
  {Zaslavsky}]{lion16}
{\sc {Lion}, S., {Alexandrova}, O. \& {Zaslavsky}, A.} 2016 {Coherent Events
  and Spectral Shape at Ion Kinetic Scales in the Fast Solar Wind Turbulence}.
  {\em The Astrophysical Journal\/} {\bf 824}, 47.

\bibitem[Matthews(1994)]{matthews94}
{\sc Matthews, Alan~P} 1994 Current advance method and cyclic leapfrog for 2d
  multispecies hybrid plasma simulations. {\em Journal of Computational
  Physics\/} {\bf 112}~(1), 102--116.

\bibitem[Narita {\em et~al.\/}(2010)Narita, Glassmeier, Sahraoui \&
  Goldstein]{narita10}
{\sc Narita, Y, Glassmeier, K-H, Sahraoui, F \& Goldstein, ML} 2010 Wave-vector
  dependence of magnetic-turbulence spectra in the solar wind. {\em Physical
  review letters\/} {\bf 104}~(17), 171101.

\bibitem[{Osman} \& {Horbury}(2009)]{osman09}
{\sc {Osman}, K.~T. \& {Horbury}, T.~S.} 2009 {Quantitative estimates of the
  slab and 2-D power in solar wind turbulence using multispacecraft data}. {\em
  Journal of Geophysical Research (Space Physics)\/} {\bf 114}, A06103.

\bibitem[Park {\em et~al.\/}(1999)Park, Belova, Fu, Tang, Strauss \&
  Sugiyama]{park99}
{\sc Park, Wonchull, Belova, EV, Fu, GY, Tang, XZ, Strauss, HR \& Sugiyama, LE}
  1999 Plasma simulation studies using multilevel physics models. {\em Physics
  of Plasmas (1994-present)\/} {\bf 6}~(5), 1796--1803.

\bibitem[Podesta(2010)]{podesta10}
{\sc Podesta, JJ} 2010 Transient growth in stable linearized vlasov--maxwell
  plasmas. {\em Physics of Plasmas (1994-present)\/} {\bf 17}~(12), 122101.

\bibitem[Podesta(2012)]{podesta12}
{\sc Podesta, John~J} 2012 The need to consider ion bernstein waves as a
  dissipation channel of solar wind turbulence. {\em Journal of Geophysical
  Research: Space Physics (1978--2012)\/} {\bf 117}~(A7).

\bibitem[Ratushnaya \& Samtaney(2014)]{ratushnaya14}
{\sc Ratushnaya, Valeria \& Samtaney, Ravi} 2014 Non-modal stability analysis
  and transient growth in a magnetized vlasov plasma. {\em EPL (Europhysics
  Letters)\/} {\bf 108}~(5), 55001.

\bibitem[Sahraoui {\em et~al.\/}(2012)Sahraoui, Belmont \&
  Goldstein]{sahraoui12}
{\sc Sahraoui, Fouad, Belmont, G{\'e}rard \& Goldstein, ML} 2012 New insight
  into short-wavelength solar wind fluctuations from vlasov theory. {\em The
  Astrophysical Journal\/} {\bf 748}~(2), 100.

\bibitem[Sahraoui {\em et~al.\/}(2009)Sahraoui, Goldstein, Robert \&
  Khotyaintsev]{sahraoui09}
{\sc Sahraoui, F, Goldstein, ML, Robert, P \& Khotyaintsev, Yu~V} 2009 Evidence
  of a cascade and dissipation of solar-wind turbulence at the electron
  gyroscale. {\em Physical review letters\/} {\bf 102}~(23), 231102.

\bibitem[Sahraoui {\em et~al.\/}(2010)Sahraoui, Goldstein, Belmont, Canu \&
  Rezeau]{sahraoui10}
{\sc Sahraoui, Fouad, Goldstein, Melvyn~L, Belmont, G{\'e}rard, Canu, P \&
  Rezeau, Laurence} 2010 Three dimensional anisotropic k spectra of turbulence
  at subproton scales in the solar wind. {\em Physical review letters\/} {\bf
  105}~(13), 131101.

\bibitem[Salem {\em et~al.\/}(2012)Salem, Howes, Sundkvist, Bale, Chaston, Chen
  \& Mozer]{salem12}
{\sc Salem, CS, Howes, GG, Sundkvist, D, Bale, SD, Chaston, CC, Chen, CHK \&
  Mozer, FS} 2012 Identification of kinetic alfv{\'e}n wave turbulence in the
  solar wind. {\em The Astrophysical Journal Letters\/} {\bf 745}~(1), L9.

\bibitem[Schekochihin {\em et~al.\/}(2009)Schekochihin, Cowley, Dorland,
  Hammett, Howes, Quataert \& Tatsuno]{schekochihin09}
{\sc Schekochihin, AA, Cowley, SC, Dorland, W, Hammett, GW, Howes, GG,
  Quataert, E \& Tatsuno, T} 2009 Astrophysical gyrokinetics: kinetic and fluid
  turbulent cascades in magnetized weakly collisional plasmas. {\em The
  Astrophysical Journal Supplement Series\/} {\bf 182}~(1), 310.

\bibitem[Servidio {\em et~al.\/}(2012)Servidio, Valentini, Califano \&
  Veltri]{servidio12}
{\sc Servidio, S, Valentini, F, Califano, F \& Veltri, P} 2012 Local kinetic
  effects in two-dimensional plasma turbulence. {\em Physical review letters\/}
  {\bf 108}~(4), 045001.

\bibitem[Stix(1962)]{stix62}
{\sc Stix, Thomas~Howard} 1962 The theory of plasma waves. {\em The Theory of
  Plasma Waves, New York: McGraw-Hill, 1962\/} {\bf 1}.

\bibitem[Swanson(2012)]{swanson12}
{\sc Swanson, Donald~Gary} 2012 {\em Plasma waves\/}. Elsevier.

\bibitem[TenBarge {\em et~al.\/}(2013)TenBarge, Howes \& Dorland]{tenbarge13}
{\sc TenBarge, JM, Howes, GG \& Dorland, W} 2013 Collisionless damping at
  electron scales in solar wind turbulence. {\em The Astrophysical Journal\/}
  {\bf 774}~(2), 139.

\bibitem[Trefethen \& Embree(2005)]{trefethen05}
{\sc Trefethen, Lloyd~Nicholas \& Embree, Mark} 2005 {\em Spectra and
  pseudospectra: the behavior of nonnormal matrices and operators\/}. Princeton
  University Press.

\bibitem[Tronci \& Camporeale(2015)]{tronci15}
{\sc Tronci, Cesare \& Camporeale, Enrico} 2015 Neutral vlasov kinetic theory
  of magnetized plasmas. {\em Physics of Plasmas (1994-present)\/} {\bf
  22}~(2), 020704.

\bibitem[Tronci {\em et~al.\/}(2014)Tronci, Tassi, Camporeale \&
  Morrison]{tronci14}
{\sc Tronci, Cesare, Tassi, Emanuele, Camporeale, Enrico \& Morrison, Philip~J}
  2014 Hybrid vlasov-mhd models: Hamiltonian vs. non-hamiltonian. {\em Plasma
  Physics and Controlled Fusion\/} {\bf 56}~(9), 095008.

\bibitem[Valentini {\em et~al.\/}(2010)Valentini, Califano \&
  Veltri]{valentini10}
{\sc Valentini, F, Califano, F \& Veltri, P} 2010 Two-dimensional kinetic
  turbulence in the solar wind. {\em Physical review letters\/} {\bf 104}~(20),
  205002.

\bibitem[Valentini {\em et~al.\/}(2007)Valentini, Tr{\'a}vn{\'\i}{\v{c}}ek,
  Califano, Hellinger \& Mangeney]{valentini07}
{\sc Valentini, F, Tr{\'a}vn{\'\i}{\v{c}}ek, P, Califano, Francesco, Hellinger,
  Petr \& Mangeney, Andr{\'e}} 2007 A hybrid-vlasov model based on the current
  advance method for the simulation of collisionless magnetized plasma. {\em
  Journal of Computational Physics\/} {\bf 225}~(1), 753--770.

\bibitem[V{\'a}sconez {\em et~al.\/}(2014)V{\'a}sconez, Valentini, Camporeale
  \& Veltri]{vasconez14}
{\sc V{\'a}sconez, CL, Valentini, F, Camporeale, E \& Veltri, P} 2014 Vlasov
  simulations of kinetic alfv{\'e}n waves at proton kinetic scales. {\em
  Physics of Plasmas (1994-present)\/} {\bf 21}~(11), 112107.

\bibitem[Waltz {\em et~al.\/}(1997)Waltz, Staebler, Dorland, Hammett,
  Kotschenreuther \& Konings]{waltz97}
{\sc Waltz, RE, Staebler, GM, Dorland, W, Hammett, GW, Kotschenreuther, M \&
  Konings, JA} 1997 A gyro-landau-fluid transport model. {\em Physics of
  Plasmas (1994-present)\/} {\bf 4}~(7), 2482--2496.

\bibitem[Zhao {\em et~al.\/}(2014)Zhao, Voitenko, Yu, Lu \& Wu]{zhao14}
{\sc Zhao, JS, Voitenko, Y, Yu, MY, Lu, JY \& Wu, DJ} 2014 Properties of
  short-wavelength oblique alfv{\'e}n and slow waves. {\em The Astrophysical
  Journal\/} {\bf 793}~(2), 107.

\end{thebibliography}

\end{document}